\documentclass[journal]{IEEEtran}

\usepackage[noadjust]{cite}
\usepackage{amsmath}
\usepackage{subfigure} 
\usepackage{graphicx}
\usepackage{adjustbox}
\usepackage{subfig}
\usepackage{tabularx}
\usepackage{array,multirow}
\usepackage{makecell}
\usepackage{adjustbox}
\usepackage{tablefootnote}
\usepackage{url}
\usepackage{cite}
\usepackage{subfigure}
\usepackage{amsmath,amssymb,amsfonts}
\usepackage{algorithmic}
\usepackage{graphicx}
\usepackage{textcomp}
\usepackage{booktabs}
\usepackage{adjustbox}
\usepackage{kotex} 
\usepackage{booktabs}
\usepackage{array}
\usepackage{tabularx}
\usepackage[font=small,labelfont=bf]{caption}

\newcolumntype{C}[1]{>{\centering\arraybackslash}p{#1}}
\newcolumntype{?}{!{\vrule width 0.8pt}}
 
\def\BibTeX{{\rm B\kern-.05em{\sc i\kern-.025em b}\kern-.08em
    T\kern-.1667em\lower.7ex\hbox{E}\kern-.125emX}}
\ifCLASSINFOpdf

\begin{document}

\title{Convolutional Recurrent Reconstructive Network for Spatiotemporal Anomaly Detection\\in Solder Paste Inspection}

\author{Yong-Ho Yoo, Ue-Hwan Kim, and Jong-Hwan Kim,~\IEEEmembership{Fellow,~IEEE}
\thanks{This work was supported by the Industrial Strategic Technology Development Program (10077589, Machine Learning Based SMT Process Optimization System Development) funded By the Ministry of Trade, Industry \& Energy (MOTIE, Korea).}

\thanks{Y.-H. Yoo is with the School of Electrical Engineering, KAIST, Daejeon, 34141, Republic of Korea, also with Koh Young Technology, Inc., Yongin-si, Gyeonggi-do, 16864, Republic of Korea (e-mail: yh.yoo@kohyoung.com).

U.-H. Kim, and J.-H. Kim are with the School of Electrical Engineering, KAIST, Daejeon, 34141, Republic of Korea (e-mail: {\{uhkim, johkim\}@rit.kaist.ac.kr}). 
}}
 
\markboth{IEEE Transactions on Cybernetics}%
{Shell \MakeLowercase{\textit{et al.}}: Bare Demo of IEEEtran.cls for IEEE Journals}
%



\maketitle

\begin{abstract}
Surface mount technology (SMT) is a process for producing printed circuit boards. Solder paste printer (SPP), package mounter, and solder reflow oven are used for SMT. The board on which the solder paste is deposited from the SPP is monitored by solder paste inspector (SPI). If SPP malfunctions due to the printer defects, the SPP produces defective products, and then abnormal patterns are detected by SPI. In this paper, we propose a convolutional recurrent reconstructive network (CRRN), which decomposes the anomaly patterns generated by the printer defects, from SPI data. CRRN learns only normal data and detects anomaly pattern through reconstruction error. CRRN consists of a spatial encoder (S-Encoder), a spatiotemporal encoder and decoder (ST-Encoder-Decoder), and a spatial decoder (S-Decoder). The ST-Encoder-Decoder consists of multiple convolutional spatiotemporal memories (CSTMs) with ST-Attention mechanism. CSTM is developed to extract spatiotemporal patterns efficiently. Additionally, a spatiotemporal attention (ST-Attention) mechanism is designed to facilitate transmitting information from the ST-Encoder to the ST-Decoder, which can solve the long-term dependency problem. We demonstrate the proposed CRRN outperforms the other conventional models in anomaly detection. Moreover, we show the discriminative power of the anomaly map decomposed by the proposed CRRN through the printer defect classification. 
\end{abstract}

\begin{IEEEkeywords}
Attention mechanism, convolutional LSTM, one-class anomaly detection, recurrent auto-encoder, spatiotemporal data, surface mount technology.
\end{IEEEkeywords}

%
\IEEEpeerreviewmaketitle

\section{Introduction}
\label{sec:introduction}
Smart factories represent fully connected and self-optimizing systems, which evolve from traditional automation factories. Smart factories collect the state data of sub-systems using IoT sensors and control sub-systems analyzing the collected data in real time. One of key technologies in smart factories is Surface Mount Technology (SMT) for automatic Printed Circuit Board (PCB) production. SMT in general consists of five steps: 1) solder paste printing, 2) solder paste inspection, 3) mounting chips, 4) chip alignment inspection and 5) reflow oven to attach chips to solder pastes. The process alternates between a task step and an inspection step, which is to guarantee a high throughput.

According to a recent analysis, 50-70\% of PCB defects occur during the solder printing step \cite{huang2011solder}. The solder paste printer (SPP) causes defects due to two factors: randomness and defects in the printer itself. SPI detects the PCB defects in advance and prevents a decrease in the throughput. SPI inspects each PCB by measuring the volume of the solder paste on each PCB pad. If the deposited solder paste is excessive, two or more pads are connected through the solder paste causing solder bridging. Conversely, if the deposited solder paste is insufficient, the reliability of the solder joint deteriorates or electrical opens could appear. Therefore, SPP is designed to deposit solder paste on each pad within the specification limit of the solder paste volume. If the deposited volume of the pad is out of the specification limit, SPI judges the pad as an excessive or insufficient one.

Traditionally, the decision thresholds for excessiveness and insufficiency are determined with the assumption that the deposited volumes on the pads with the same aperture shape follow a normal distribution. With the assumption, SPI groups the deposited pads by the same aperture shape, and calculates the mean and variance of the volumes per each group. Then, the pads whose solder paste volumes are far from the mean are regarded as anomaly pads. This approach, however, can hardly identify the PCB defects caused by the printer defects. The printer malfunction increases the ratio of the anomaly pads, which makes the normal distribution biased. 

To overcome the above-mentioned limitation, we propose a Convolutional Recurrent Reconstructive Network (CRRN) that detects anomaly pads without the normal distribution assumption. CRRN, as a type of convolutional recurrent autoencoder (CRAE), consists of a spatial encoder (S-Encoder), a spatiotemporal encoder and decoder (ST-Encoder-Decoder), and a spatial decoder (S-Decoder). By training CRRN using only normal data, CRRN can reconstructs normal data even if it receives anomaly data. Thus, the anomaly map can be decomposed from the reconstruction error between the reconstructed data and the input data. 
%

Two main features proposed in this paper characterize CRRN from conventional CRAEs: Convolutional Spatiotemporal Memory (CSTM) and spatiotemporal attention (ST-Attention) mechanism. First, CSTM helps CRRN to utilize both the spatial and temporal information whereas conventional CRAEs exploits only temporal information. A CSTM cell directly delivers spatial information to other CSTM cells without incrementing the number of parameters. Next, the ST-Attention mechanism effectively handles long-term dependency problem by transmitting the ST-Attention map extracted in the encoder to the decoder in CRRN, which acts as a shortcut path between the encoder and decoder. 

The main contributions of our work are as follows.
\begin{enumerate}
\item We develop CSTM that boosts the performance of the proposed CRRN by exploiting both spatial and temporal information.
\item We design the ST-Attention mechanism to deal with long-term dependency.
\item We formulate the spatiotemporal anomaly detection of the SPI data using CRRN. 
\item We conduct a series of experiments and analyze the experiment results thoroughly to verify the performance of CRRN in anomaly detection.
\end{enumerate}

The rest of the paper is structured as follows. In Section II, we discuss previous works related to CRRN. Section III delineates the characteristics of SPI data. Section IV presents the proposed CRRN and Section V verifies the performance of CRRN in anomaly detection. Concluding remarks follow in Section VI.


\section{Related Works}
\subsection{Anomaly Detection}
Anomaly detection can be classified into spatial anomaly detection, temporal anomaly detection, and spatiotemporal anomaly detection depending on the data types it handles. First, spatial anomaly detection identifies anomaly regions in the spatial data such as images. Recent studies for the spatial anomaly detection mainly utilize unsupervised learning algorithms such as the generative adversarial network \cite{schlegl2017unsupervised}, and the convolutional auto-encoder \cite{baur2018deep, mei2018automatic, haselmann2018anomaly, soukup2014convolutional,masci2012steel}. Application areas of spatial anomaly detection include medicine and manufacturing. In the medical field, anomaly detection highlights the disease regions in medical images \cite{schlegl2017unsupervised, baur2018deep}. Anomaly detection in manufacturing identifies defective parts of products using the product images \cite{soukup2014convolutional,masci2012steel}. 


Next, temporal anomaly detection identifies the anomaly regions in the time-varying data. One-class SVM \cite{khreich2017anomaly} and recurrent auto-encoder (RAE) \cite{malhotra2015long, jurgovsky2018sequence, singh2015intrusion, mirza2018computer, chauhan2015anomaly} are main algorithms used for the temporal anomaly detection. In the finance field, researchers have studied to minimize the damage caused by unexpected patterns. Unusual power consumption can be predicted by analyzing the current power consumption \cite{malhotra2015long}. A crime such as credit card fraud can be detected in advance \cite{jurgovsky2018sequence}. In security, there are studies on intrusion detection using system logs or temporal network signals \cite{khreich2017anomaly, singh2015intrusion, mirza2018computer}. In health-care, modeling biological data such as Electrocardiography (ECG) signals can be utilized to detect the abnormal signals of the body in advance. \cite{chauhan2015anomaly}.  

Spatiotemporal anomaly detection identifies the anomaly regions in the spationtemporal data. Recently, the combinations of RNN with CNN have been widely applied to extract both of the spatial and temporal patterns \cite{xingjian2015convolutional,medel2016anomaly,wang2017predrnn}. Intelligent traffic system utilizes the video streams for preventing accidents or predicting traffic jams \cite{huang2018traffic}. Video-surveillance monitors temporal images and then detects the abnormal events \cite{zhao2017spatio, luo2017remembering, xu2015learning}. In this paper, we focus on the spatiotemporal anomaly detection of SPI data obtained during SMT process. 

\subsection{SMT Optimization}
The quality of the soldering can be enhanced by optimizing the key factors \cite{khader2018stencil,khader2017stencil}: squeegee pressure, squeegee speed, and stencil cleaning interval. By analyzing the trend of soldering, the stencil cleaning interval can be automatically optimized \cite{chang2019implementation, wang2018recurrent}. Similarly, the optimization of SPP parameters improves the quality of the end-products \cite{gopal2006optimization}. In addition, there are studies diagnosing the quality of soldering and identifying soldering defects \cite{hui2009solder, wu2017solder, benedek2011detection}. Their studies are to diagnose soldering states of a single PCB. However, there are limits to applying to the SPP diagnosis, since the patterns of the SPP defects may appear over multiple PCBs. In this paper, we propose the CRRN that can decompose anomaly maps using SPI data of multiple PCBs to diagnose the SPP defects. In particular, we decompose the anomaly map caused by the printer defects. 

\section{Characteristics of SPI Data}

\begin{figure}
    \centering
    \includegraphics[width=0.65\linewidth]{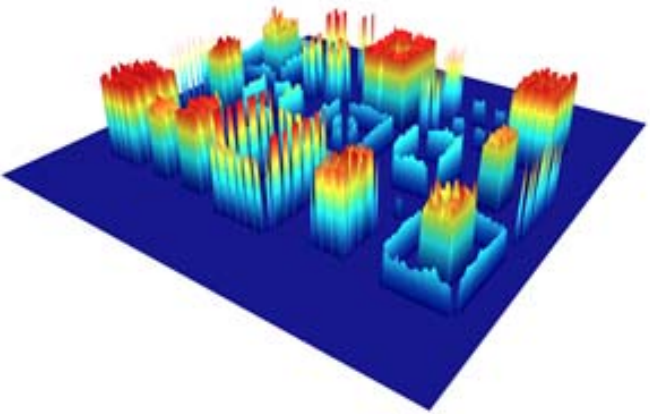}
    \caption{Visualization of the 3D SPI data.}
    \label{fig:3dspi}
\end{figure}
\begin{figure}
	\centering	   	
    \subfigure[]{\includegraphics[width=\linewidth]{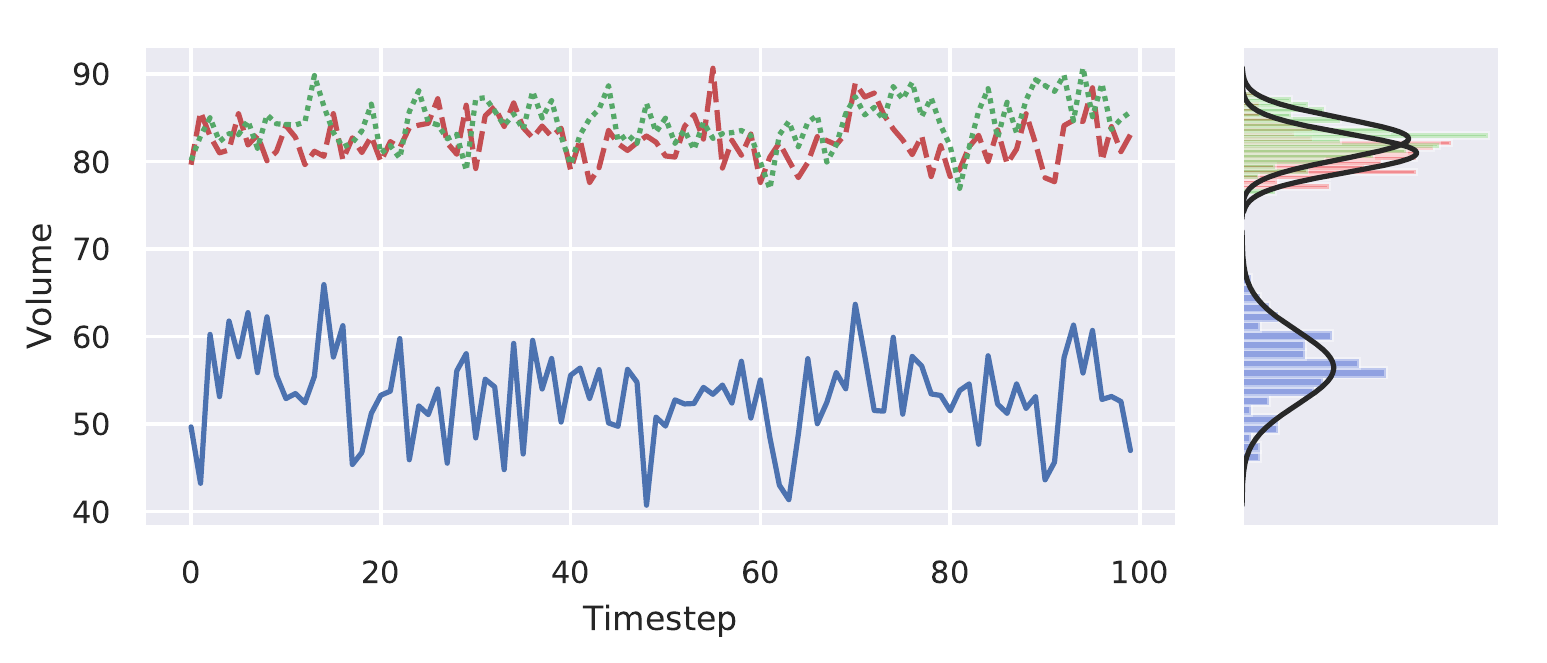}}    
	\subfigure[]{\includegraphics[width=\linewidth]{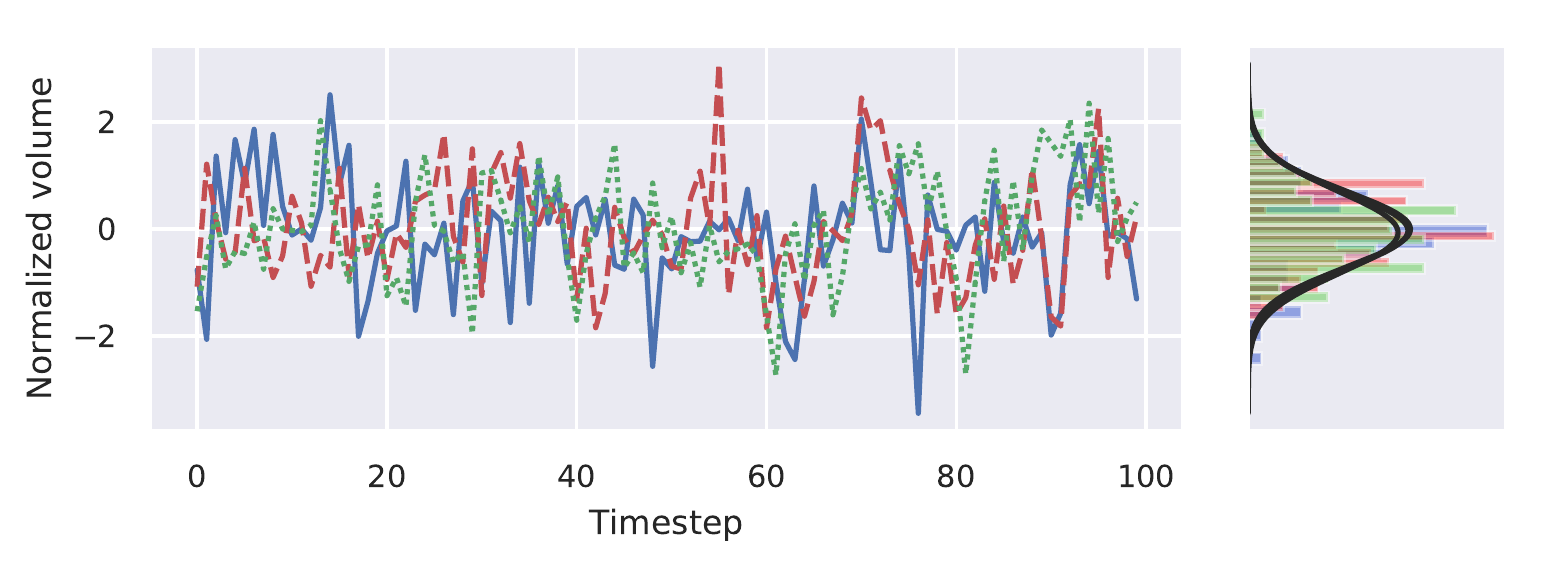}}
	\subfigure[]{\includegraphics[width=\linewidth]{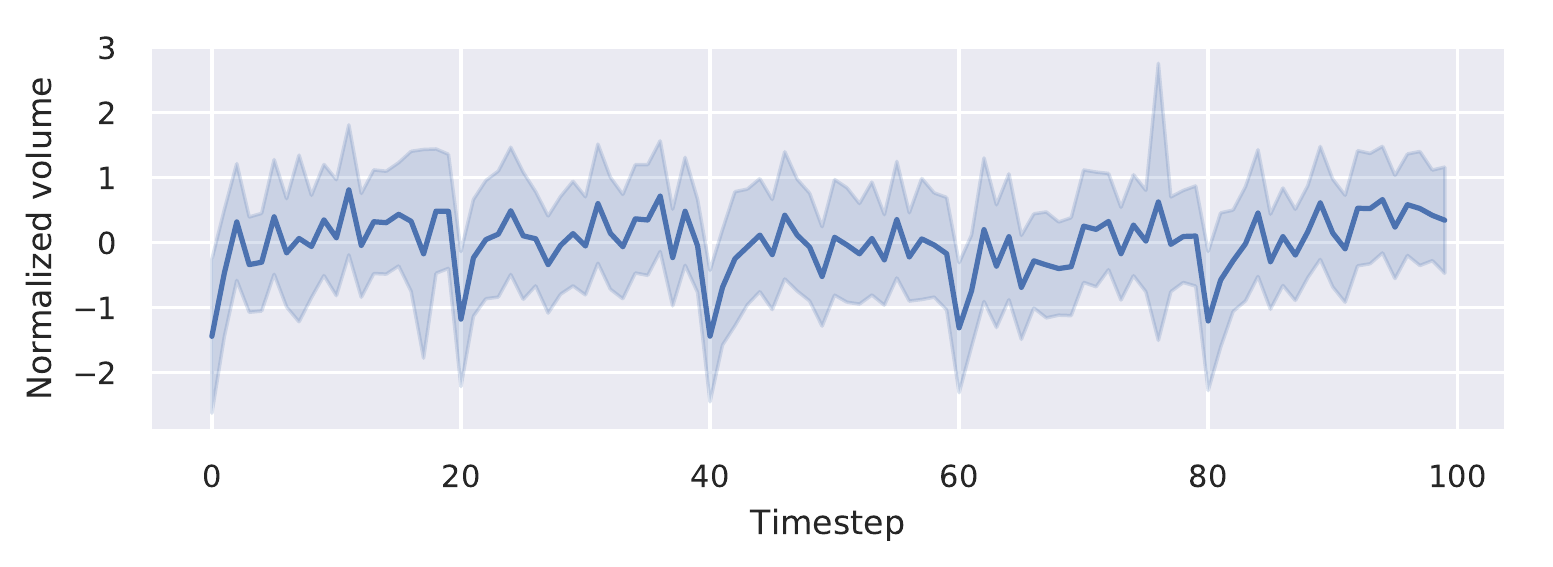}}
    \caption{(a) The profiles of the three arbitrarily selected solder paste volumes over timestep, (b) the profile of the normalized volume over timestep, and (c) the profile of the average normalized volume over timestep. Timestep indicates the number in the order that the PCB products are produced.}
    \label{fig:vol} 
\end{figure}

Several packages, mounted on the board, are fixed to the board with the deposited solder paste. Thus, the position of the deposited solder paste should match the position of the metal pad in the package. As shown in Fig. \ref{fig:3dspi}, the SPI data, measuring the volume of the solder paste, contains a spatial pattern due to the different spatial locations of the packages and various shapes of pads. The solder paste volume of each pad depends on the package type. Moreover, even in the same package, the solder paste volume of each pad varies according to the relative position in the package.


Also, SPI data display a temporal pattern in addition to the spatial pattern. The temporal pattern originates from two factors: the stencil cleaning interval of SPP and the two squeegee blades of SPP. Firstly, SPP regularly cleans up the residual solder paste of the stencil to prevent anomaly pads. During mass production of PCBs using SPP, residual solder paste accumulates beneath the stencil and inside the stencil apertures. The residual solder paste beneath the stencil causes a solder smudging problem and the residual solder paste inside the stencil apertures causes an aperture blockage problem. To resolve the problems, the stencil cleaning is periodically performed to remove the remaining solder paste and the periodic cleaning generates the temporal pattern of the SPI data.

Second, SPP utilizes two squeegee blades to maximize printing efficiency. One blade deposits the solder paste by rolling solder paste on the stencil in the forward direction, while the other blade deposits the solder paste in the backward direction. The two blades perform the deposition of solder paste in a regular manner and show two distinctive characteristics. As a result, the SPI data display temporal patterns along the two directions of the blades. 


Fig. \ref{fig:vol}(a) shows the volume of three arbitrarily selected pads over timestep that indicates the order that the PCB products are produced. The graph on the right side in the figure shows the distributions of the solder paste volume for each pad. Each distribution shows different distributions due to different spatial features such as the pad location and the shape. The normalization of the solder paste volume distribution eliminates the spatial pattern as shown in Fig. \ref{fig:vol}(b). Fig. \ref{fig:vol}(c) plots the average of normalized distributions over timestep. As shown in Fig. \ref{fig:vol}(c), the average of normalized distributions has periodicity, and the period corresponds to the stencil cleaning interval. Moreover, the average of normalized distributions moves up and down in order due to the two directions of squeegee blades in SPP. 
\begin{figure*}
\centering
\includegraphics[width=0.7\linewidth]{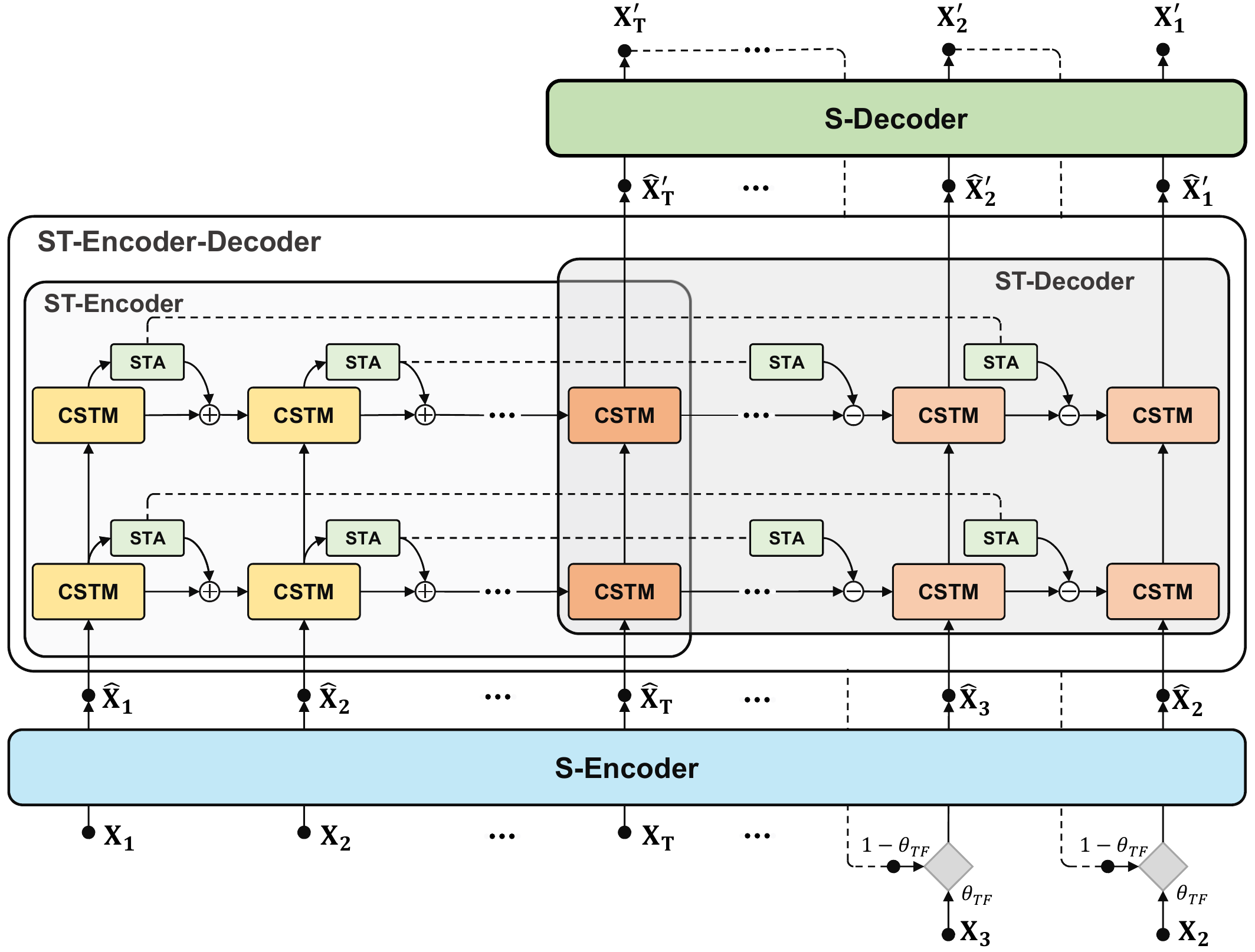}
\caption{The overall structure of the proposed CRRN model. STA stands for ST-Attention.}
\label{fig:crrn} 
\end{figure*}
\section{Convolutional Recurrent Reconstructive Network}
In this section, we describe the proposed Convolutional Recurrent Reconstructive Network (CRRN) for anomaly detection in spatiotemporal data. Fig. \ref{fig:crrn} displays the overall architecture of CRRN. CRRN is a type of an encoder-decoder model. The encoder consists of a spatial encoder (S-Encoder) that extracts spatial features from the spatiotemporal input, and a spatiotemporal encoder (ST-Encoder) that extracts spatiotemporal features from a sequence of the spatial features. Similarly, the decoder of CRRN consists of an S-Encoder, a spatiotemporal decoder (ST-Decoder) that decodes spatial features of each timestep, and a spatial decoder (S-Decoder) that reconstructs the original data. 
\begin{figure}
	\centering	   	
    \subfigure[]{\includegraphics[width=0.96\linewidth]{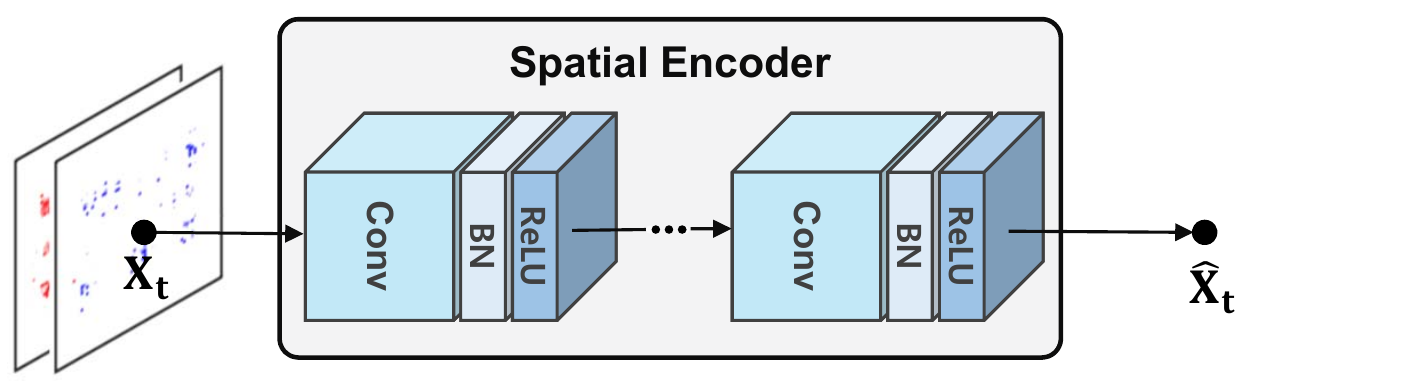}}    
	\subfigure[]{\includegraphics[width=0.96\linewidth]{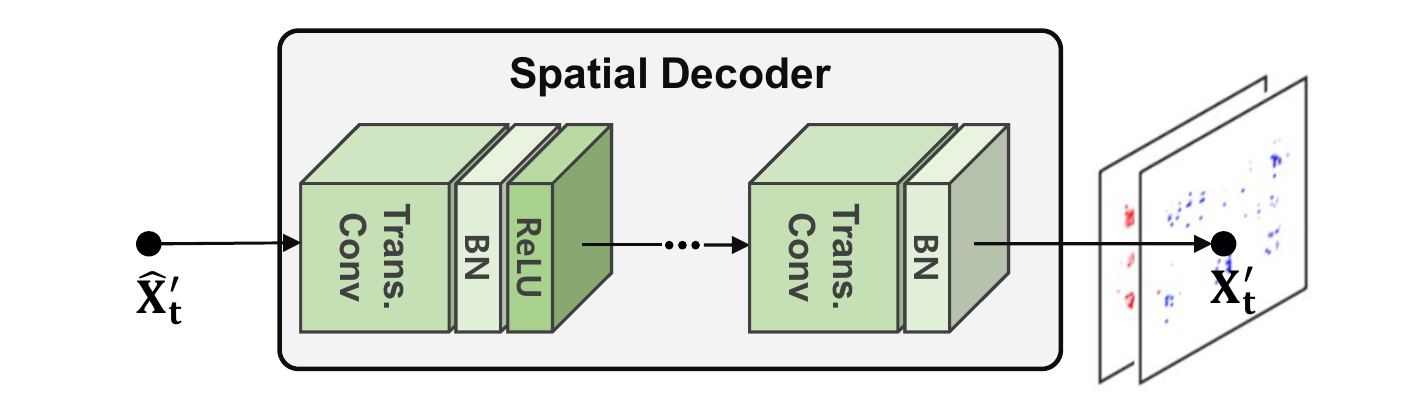}}
        \caption{The structures of (a) the spatial encoder, and (b) the spatial decoder.} 
    \label{fig:s_enc_dec} 
\end{figure}
\subsection{Spatial Encoder and Decoder}
The S-Encoder shown in Fig. \ref{fig:s_enc_dec}(a), extracts the spatial feature, $\hat{X}_t \in \mathbb{R}^{N_c \times N_h \times N_w}$ from the original spatial data, $X_t \in \mathbb{R}^{N_c^{in} \times N_h^{in} \times N_w^{in}}$, where $N_c$ ($N_c^{in}$), $N_h$ ($N_h^{in}$), and $N_w$ ($N_w^{in}$) respectively denote the numbers of channel, height, and width of $\hat{X}_t$ ($X_t$). The S-Encoder contains multiple modules and each module consists of a convolution layer, a batch normalization layer, and a ReLU activation layer. The convolution layer converts high dimension feature maps into low dimension feature maps. The batch normalization prevents the covariance shift by normalizing the values of each layer \cite{ioffe2015batch}. The ReLU activation layer imposes non-linearity to enhance the modeling capability of CRRN.

The S-Decoder shown in Fig. \ref{fig:s_enc_dec}(b), operates in the reverse manner compared to the S-Encoder. The S-Decoder generates the reconstructed spatial data, $X_t^{'}\in \mathbb{R}^{N_c^{out} \times N_h^{out} \times N_w^{out}}$ from the low-dimensional spatial feature, $\hat{X}_t^{'} \in \mathbb{R}^{N_c \times N_h \times N_w}$. The dimension of $X_t$ is equal to that of $X_t^{'}$. The S-Decoder also consists of multiple modules. Unlike the S-Encoder, the S-Decoder uses a deconvolutional operation known as the transposed convolution instead of the convolutional operation. The deconvolution reconstructs a high dimensional feature maps from a low dimension feature maps. The last module of the S-Decoder does not include the ReLU activation layer, thus ensuring that the generated output has a valid range.
%
\subsection{Spatiotemporal Encoder and Decoder}
\subsubsection{Convolutional Spatiotemporal Memory (CSTM)} 
\begin{figure}
\centering
\includegraphics[width=0.85\linewidth]{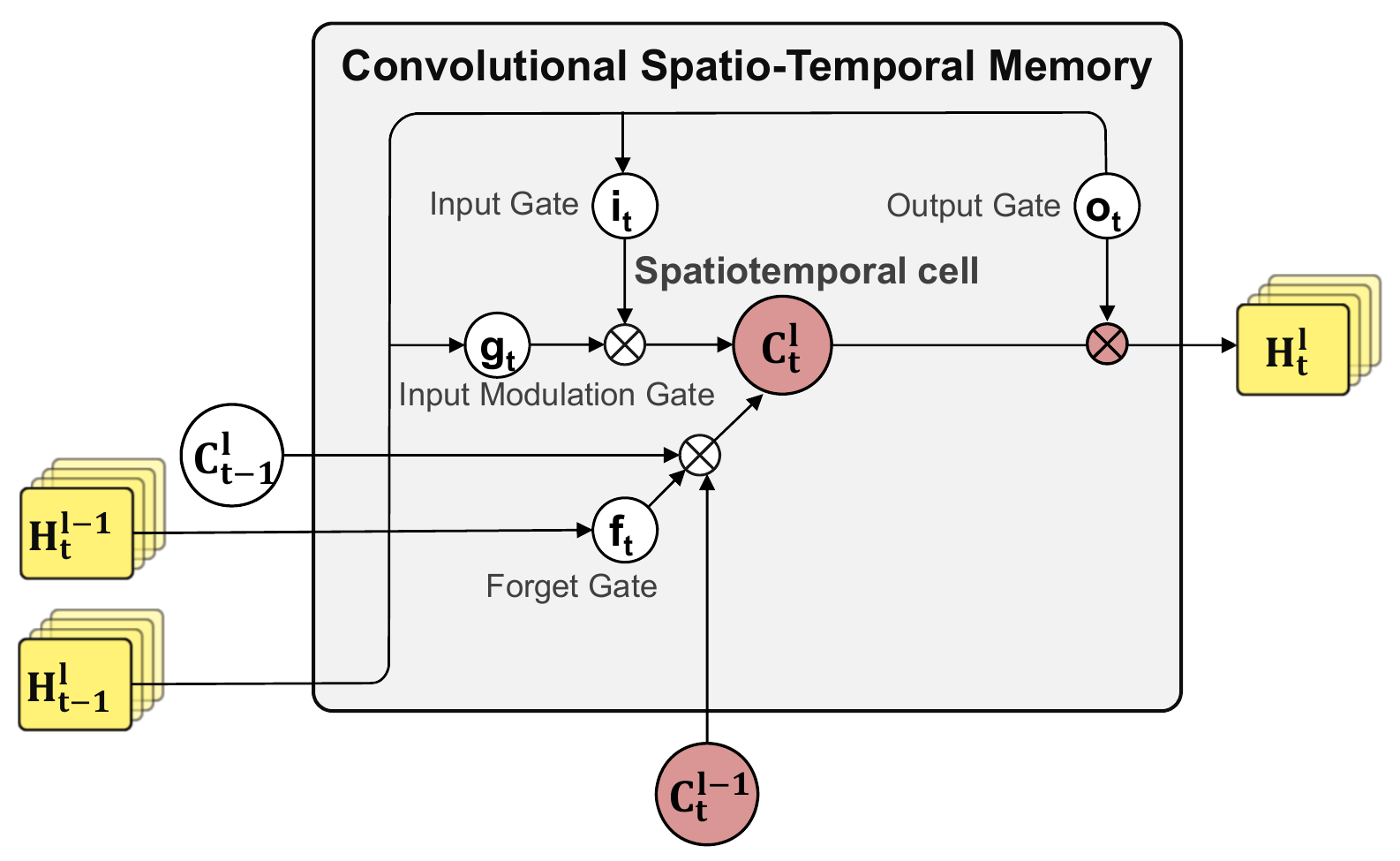}
\caption{Convolutional Spatiotemporal Memory.}
\label{fig:cstm} 
\end{figure}
We develop the proposed CSTM based on Convolutional LSTM (ConvLSTM) and Spatiotemporal LSTM (ST-LSTM). In contrast to the vanilla LSTM which can hardly extract spatial features, the convolutional operation \cite{xingjian2015convolutional} enables the ConvLSTM to extract both of the temporal and spatial information simultaneously, while using less parameters than that of the LSTM. We describe the operation of ConvLSTM for the formulation of CSTM as follows:
\begin{subequations}
\begin{gather}
i_t^l = \sigma (W_i*H_t^{l-1} + U_i * H_{t-1}^l + W_{c} \circ C_{t-1}^l), \\ 
f_t^l = \sigma (W_f*H_t^{l-1} + U_f * H_{t-1}^l + W_{f} \circ C_{t-1}^l),\\
C_t^l = f_t^l  \circ C_{t-1}^l + i_t^l \circ tanh (W_g * H_t^{l-1} + U_c * H_{t-1}^l), \\
o_t^l = \sigma (W_o*H_{t-1}^l + U_o * H_{t-1}^l+ W_{o} \circ C_{t}^l), \\
H_t^l = o_t^l \circ tanh (C_t^l)
\end{gather}
\label{eq:convlstm}%
\end{subequations}
where the superscript $l$ and $t$ denote a layer and a timestep, respectively, $i_t^l$, $f_t^l$, and $o_t^l$ denote an input gate, forget gate, and output gate, respectively, $H_t^0$ is equal to $\hat{X}_t$, $*$ is a convolutional operator and $\circ$ is a Hadamard product. The cell gate, $C_t^l$ is a function of $C_{t-1}^l$, which helps the flow of the temporal information. However, the $C_t^l$ does not rely on $C_t^{l-1}$ directly, which causes a part of the spatial information of the input feature to be lost. Spatiotemporal LSTM (ST-LSTM) is designed to facilitate the flows of the spatiotemporal information \cite{wang2017predrnn}. The ST-LSTM has another cell gate responsible for the spatial information flow, as well as the cell gate, $C_t^l$ responsible for the temporal information flow. However, the ST-LSTM requires twice as many parameters as the ConvLSTM.

Fig. \ref{fig:cstm} shows the proposed CSTM. The CSTM can effectively capture the spatiotemporal pattern while maintaining the similar number of parameters of the ConvLSTM. The CSTM replaces the cell gate, $C_t^l \in \mathbb{R}^{N_c \times N_h \times N_w}$ in (\ref{eq:convlstm}c), with the following operation: 
\begin{equation}
C_t^l = f_t^l \circ W_{1 \times 1} [C_{t-1}^l ; C_t^{l-1}] + i_t^l \circ g_t^l
\end{equation}
where the two cell gates, $C_t^{l-1}$ and $C_{t-1}^l$, each with $N_c$ channels, are concatenated in a channel-wise manner. $W_{1\times 1} \in \mathbb{R}^{N_c \times 2N_c}$ is the weight matrix of the one by one convolutional operation for reducing the number of channels back to half. The one-by-one convolution, whose kernel size, $N_k$ is 1, has the advantage of adjusting the number of channels while keeping the dimension of the feature map. 

\subsubsection{ST-Encoder and ST-Decoder} 

The ST-Encoder-Decoder part of CRRN employs the proposed ST-Encoder and ST-Decoder. For readability, we denote the hidden states of the ST-Encoder and ST-Decoder by $E_t^l \in \mathbb{R}^{N_c \times N_h \times N_w}$ and $D_t^l \in \mathbb{R}^{N_c \times N_h \times N_w}$, respectively, instead of $H_t^l$. The hidden state, $E_t^l$ in the ST-Encoder is updated as follows: 

%
\begin{equation}
E_t^l  = CSTM(E_t^{l-1},E_{t-1}^l).
\end{equation} 
The hidden state of the last timestep, $E_T^l$ is copied to the hidden state, $D_T^l$ as shown in Fig. \ref{fig:crrn}. The hidden state, $D_t^l$ in the ST-Decoder is updated as follows: 
\begin{equation}
D_t^l=CSTM(D_t^{l-1},D_{t+1}^l)
\label{eq:cstm} 
\end{equation}
where the hidden state of the top layer, $D_t^L$ is denoted by $\hat{X}_t$.
For training the ST-Encoder-Decoder model, we adopt the scheduled sampling \cite{bengio2015scheduled} for feeding inputs. The scheduled sampling probabilistically selects between the generated output of the previous time step, $X_t^{'}$ and the ground truth of the previous time step, $X_t$. The scheduled sampling is formulated as follows:
\begin{equation}
X_t^{in} = \theta_{TF}{X}_t + (1-\theta_{TF})X_t^{'}
\label{eq:tf} 
\end{equation}
where $X_t^{in}$ is the input of the model, $X_t$ and ${X}_t^{'}$ are ground truth input and generated input, respectively and $\theta_{TF} \in \{ 0, 1\}$ is a sampling mask that chooses between $X_t$ and ${X}_t^{'}$. $\theta_{TF}$ follows a Bernoulli distribution, $P(\theta_{TF})=\epsilon^{\theta_{TF}}(1-\epsilon)^{1-\theta_{TF}}$ where $\epsilon$ is the probability of selecting ${X}_t$. In the early state of learning, $\epsilon$ is set high and gradually decreased over the training epochs. The scheduled sampling technique can solve the exposure bias problem by guiding the learning process similar to the test process.

In addition, we apply a denoising auto-encoder technique to prevent CRRN from acting like an identity function. We add random noise to the inputs during the training phase. Moreover, we set a part of input elements to zero at the training phase. The denoising auto-encoder enhances the resilience of the original data without the noises \cite{vincent2008extracting}. 

\subsubsection{Spatiotemporal Attention Mechanism}
\begin{figure}
\centering
\includegraphics[width=1\linewidth]{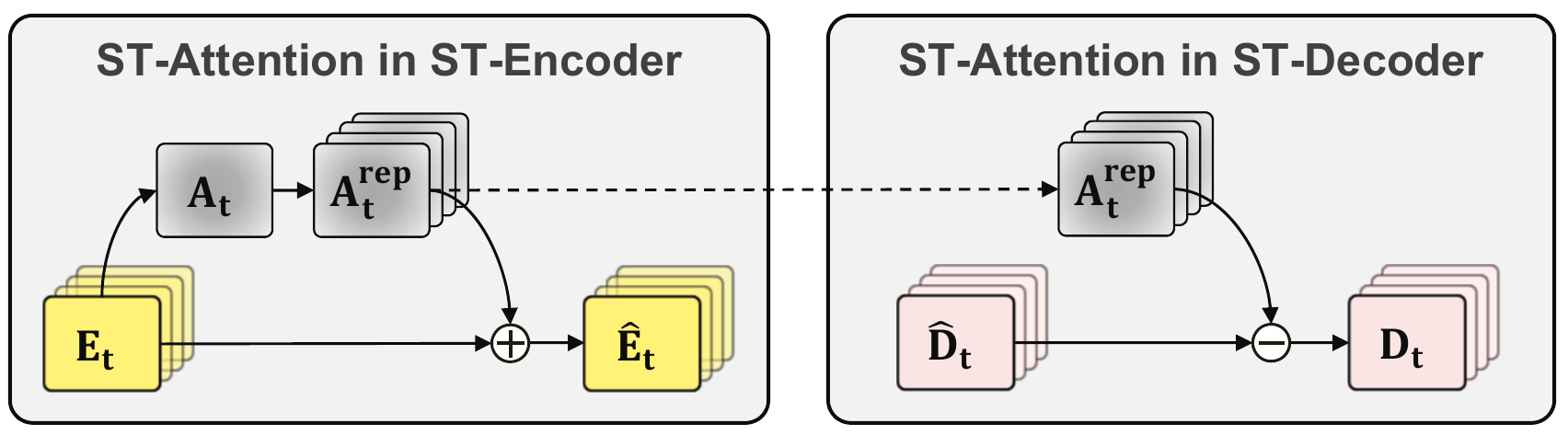}
\caption{ST-Attention mechanism in ST-Encoder and ST-Decoder.}
\label{fig:attention} 
\end{figure}
As the length of the sequence increases, the number of hidden states to pass to reach from the hidden state of the encoder to the corresponding hidden state of the decoder increases. This causes long-term dependency problem. To resolve the problem, we propose a spatiotemporal attention (ST-Attention) mechanism that directly transmits the information of the corresponding hidden states between the encoder and decoder, regardless of the sequence length. The ST-Attention mechanism extracts the ST-Attention map that represents the feature of the hidden state of the encoder, and then reuses the ST-Attention map when generating the corresponding hidden state in the decoder. Since the ST-Attention map acts as a bridge between the encoder and the decoder, the proposed ST-Attention mechanism can effectively solve the long-term dependency problem.

The procedure of the ST-Attention mechanism is shown in Fig. \ref{fig:attention}. The ST-Attention mechanism extracts the spatial feature of $E_t$ that affects the next hidden state, $E_{t+1}$. For this, we define the ST-Attention map, $A_t \in \mathbb{R}^{1 \times N_h \times N_w}$ that is applied to the hidden state, $E_t$ before transmitting to next hidden state, $E_{t+1}$ as follows: 
\begin{equation}
A_t = tanh (W_A * E_t)
\end{equation}
where $W_A \in \mathbb{R}^{1 \times N_c \times N_k \times N_k}$ represents the weight matrices of the convolutional operation and $N_k$ denotes the kernel size of the convolution. The hidden state, $E_t$ with $N_c$ channels is converted to the ST-Attention map, $A_t$ with 1 channel. $A_t$ is replicated in channel-wise manner to match the number of channels with that of $E_t$. The replicated ST-Attention map, $A_t^{rep}$ is applied to the hidden state, $E_t$ as follows: 
\begin{equation}
\hat{E}_t = E_t + A_t^{rep}.
\label{eq:appl_attn}
\end{equation}

In the decoder, the ST-Attention map, $A_t$ obtained in the encoder is applied to the hidden states in the reverse way. We subtract $A_t^{rep}$ from $\hat{D}_t$ and obtain $D_t$ as follows: 
\begin{equation}
D_t = \hat{D}_t - A_t^{rep}.
\label{eq:release_attn}
\end{equation}
We reuse the $A_t$ obtained from the encoder. Thus, $A_t$ acts as a shortcut path between the encoder and the decoder. The reason for reversing the operation of the attention map in the encoder and decoder is to use the last hidden state of the encoder, $E_T$ as the initial hidden state of the decoder, $D_T$.  

The last hidden state, $E_T$ is a feature that contains global information of the input sequence. The longer the length of the sequence, the more difficult for the fixed-size hidden state to encode the information of the input sequence due to increased contexts. If the ST-Attention mechanism is applied, the local information of each input can be efficiently transmitted through the ST-Attention map, which makes the decoder directly access the corresponding ST-Attention map when generating each output. Since local information is captured in the ST-Attention map, global information can be captured intensively in the last hidden state of the encoder, $E_t$.

\section{Experiments}
We evaluated the performance of the proposed CRRN for the anomaly detection through three experiments. Two experiments were conducted to verify the performance of the anomaly map decomposition and one experiment was conducted to classify the SPP defects using the decomposed anomaly map. For the anomaly map decomposition, we used two layers for all of the S-Encoder, S-Deocder and ST-Encoder-Decoder. The number of channels in the input and output layers were set to 2, and those of the rest layers were set to 64. The kernel size was set to $5\times5$ for all convolutional layers. Under the experimental setting, we compared the performance of the proposed CRRN with those of the statistical method and the CRAE consisting of ConvLSTMs. In addition, we compared the performance without and with the ST-Attention mechanism. We denoted CRRN and CRAE without the ST-Attention mechanisms as CRRN\textbackslash a and CRAE\textbackslash a, respectively.

In Experiment 1, we evaluated the performance for synthetic SPI data whose anomaly pads were generated by adding noises to randomly selected pads. In Experiment 2, real data obtained from the defective SPP were used to decompose the anomaly map. In addition to the two experiments, we demonstrated the discriminative power of the anomaly map generated from the CRRN for a SPP defect classification task. 

\subsection{Experiment 1} 
Fig. 7(a) shows the process of generating the synthetic SPI data including the anomaly pads. The synthetic SPI data, $X_{in}^t$ was generated by adding a randomly generated anomaly map, $\epsilon_t^{generated}$ to the normal SPI data, $X_t^{normal}$ as follows: 
\begin{align}
X_t^{in} & = X_t^{normal} + \epsilon_t^{generated} \nonumber \\
 & = X_t^{normal} + M_t^{label} \cdot M_t^{N(\mu, \sigma^2)} \nonumber \\
  & = X_t^{normal} +  M_t^{pad} \cdot M_t^{anomaly} \cdot M_t^{N(\mu, \sigma^2)}
\label{eq:1}
\end{align}
where dimensions of $X_{t}^{in}$, $\epsilon_t^{generated}$, $X_t^{normal}$, $M_t^{label}$, $M_t^{N(\mu, \sigma^2)}$, $M_t^{pad}$, and $M_t^{anomaly}$ are all $\mathbb{R}^{N_c^{in} \times N_h^{in} \times N_w^{in}}$. $\epsilon_t^{generated}$ is expressed as product of the $M_t^{label}$ and $M_t^{N(\mu, \sigma^2)}$. $M_t^{label}$ indicates whether each pad is an anomaly or not. $M_t^{label}$ is determined by the product of $M_t^{pad}$ and $M_t^{anomaly}$. $M_t^{pad}$ indicates the map where the pads actually exist in the SPI data, and $M_t^{anomaly}$ is a randomly generated binary map. Each element in $M_t^{N(\mu, \sigma^2)}$ is sampled from Gaussian noise, $N(\mu, \sigma^2)$. 

\begin{figure}
	\centering	   	
    \subfigure[]{\includegraphics[width=1\linewidth]{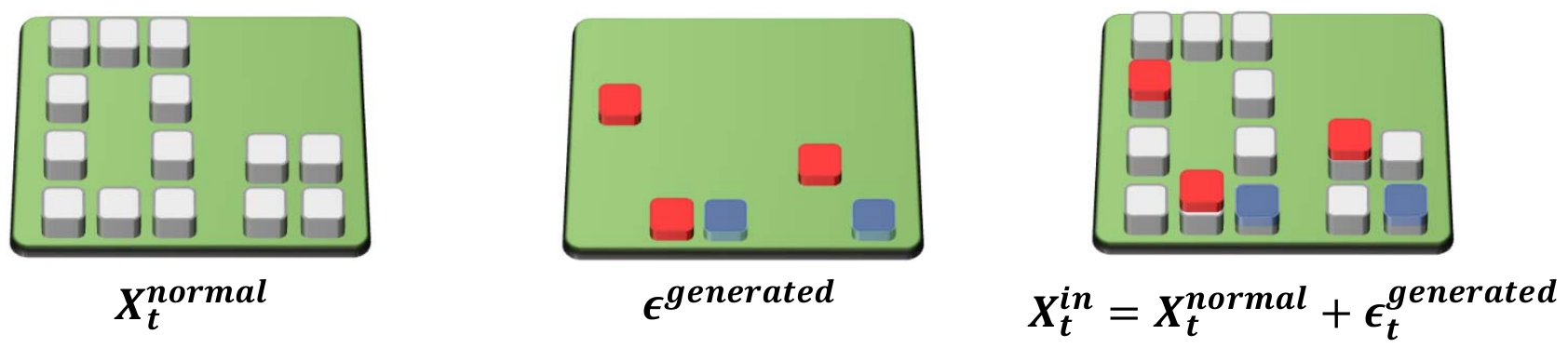}}    
        \subfigure[]{\includegraphics[width=1\linewidth]{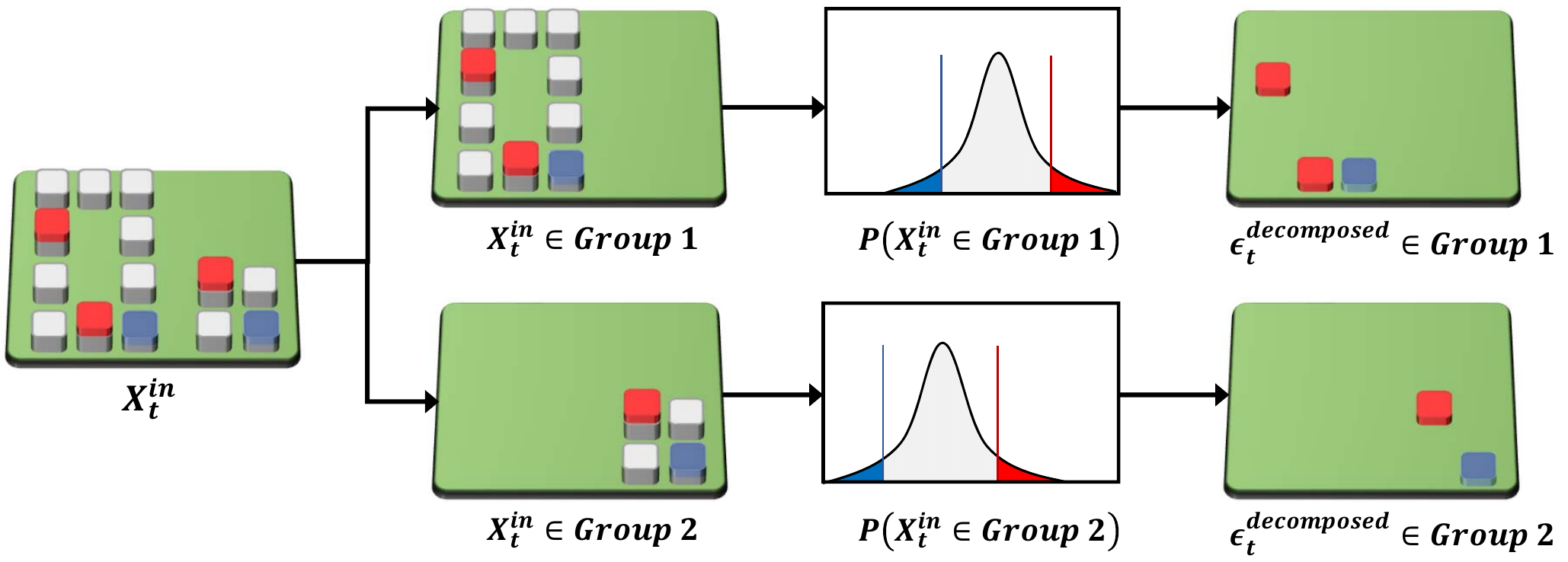}}    
	\subfigure[]{\includegraphics[width=1\linewidth]{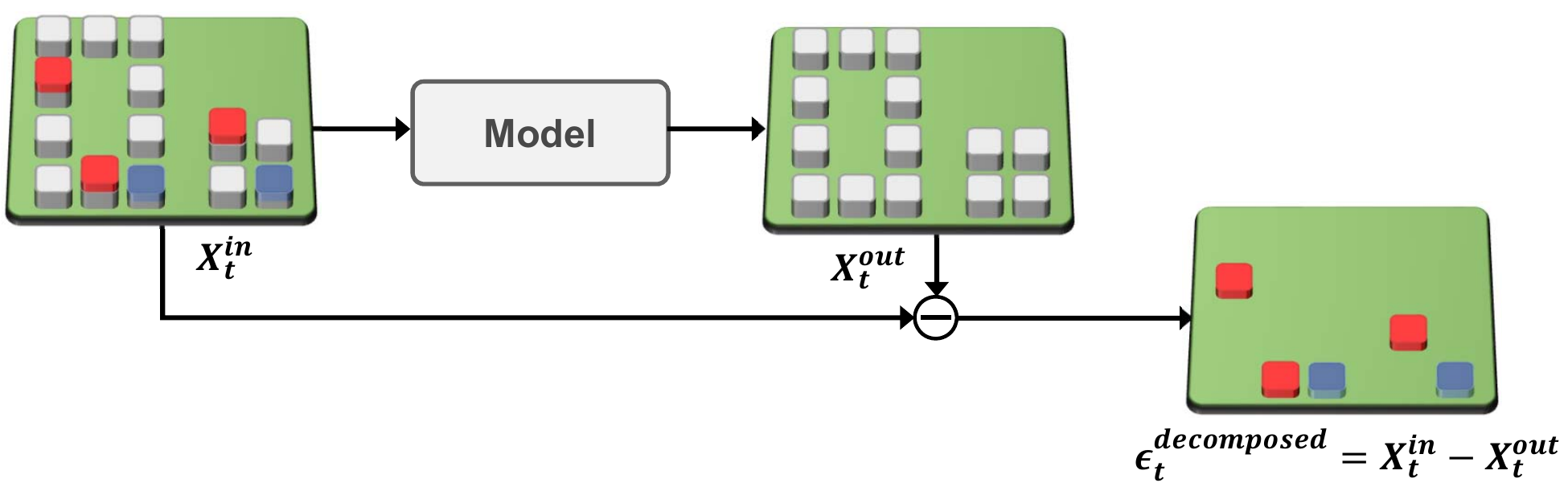}}
    \caption{Experiment 1. (a) Anomaly pad generation for evaluating the performance of anomaly detection. (b) Anomaly detection using the statistical method. (c) Anomaly detection using the deep-learning based method.} 
    \label{fig:simulation1_process} 
\end{figure}

Fig. \ref{fig:simulation1_process}(b) shows the process of detecting the anomaly pads through the statistical method. By assuming that the volumes of pads with the same shape follow the Gaussian distributions, the pads whose volumes deviate from the mean are regarded as the excessive or insufficient pads. Fig. \ref{fig:simulation1_process}(c) shows the process of detecting the anomaly pads through the deep-learning based model. Since the model is trained to reconstruct the normal data even if abnormal data are fed into the model, the reconstructed output is regarded as the normal SPI data. The decomposed anomaly map, $\epsilon_t^{decomposed}$ is the reconstruction error that is calculated by subtracting $X_t^{out}$ from $X_t^{in}$. 

In this experiment, we generated the random mask, $M_t^{N(\mu, \sigma^2)}$ with $\mu$ = $[5,-5]$ and $\sigma$ = $0.1$ to cover a broad range of the noises. In addition to generating different scales of the noises, we generated $M_t^{anomaly}$ by sweeping the anomaly ratio, $P_a=[10, 20, 30, 40, 50]\%$ to compare the performance according to the anomaly pad ratio in the SPI data. From the anomaly map, $\epsilon_t^{decomposed}$, the pads whose values are larger than the decision threshold are judged as the anomaly pads. 

\begin{figure*}
\centering
\includegraphics[width=1\linewidth]{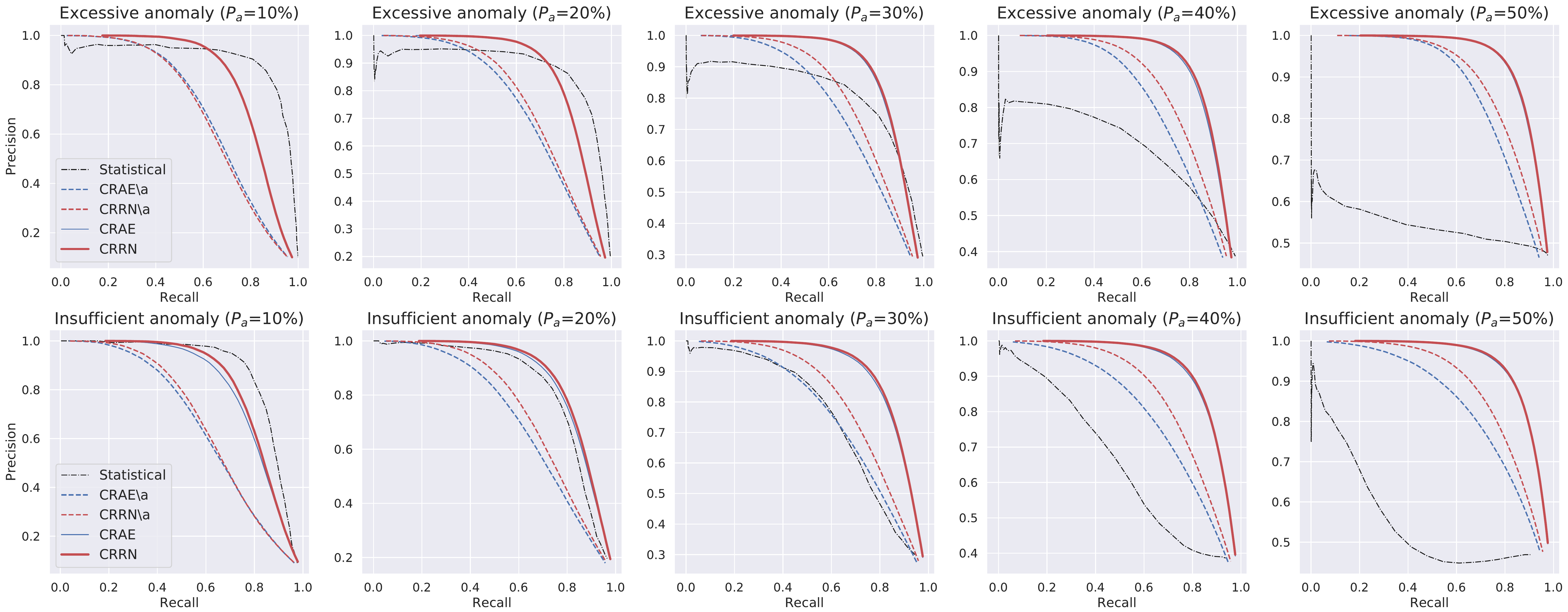}
\caption{Precision-recall curves for anomaly detection (Upper and lower graphs represent excessive and insufficient pad detection results, respectively.}
\label{fig:pr_curve} 
\end{figure*}

By sweeping over the decision thresholds, the precision-recall (PR) curves of the statistical method and deep-learning based methods (CRAE\textbackslash a, CRRN\textbackslash a, CRAE, and CRRN) are plotted in Fig. \ref{fig:pr_curve}. The sub-figures in the figure from left to right show the PR curves according to the anomaly ratio, $P_a$ ranging from $10\%$ to $50\%$. The upper and lower rows in the figure represent the PR curves of detecting the excessive and insufficient pads, respectively. When $P_a$ is $10\%$, the anomaly detection accuracy of the statistical method is similar to those of deep-learning based methods. As the anomaly ratio increases, however, the performance of the statistical method drops sharply. The reason for this is described in the following. 

\begin{figure*}
    \centering
    \includegraphics[width=1\linewidth]{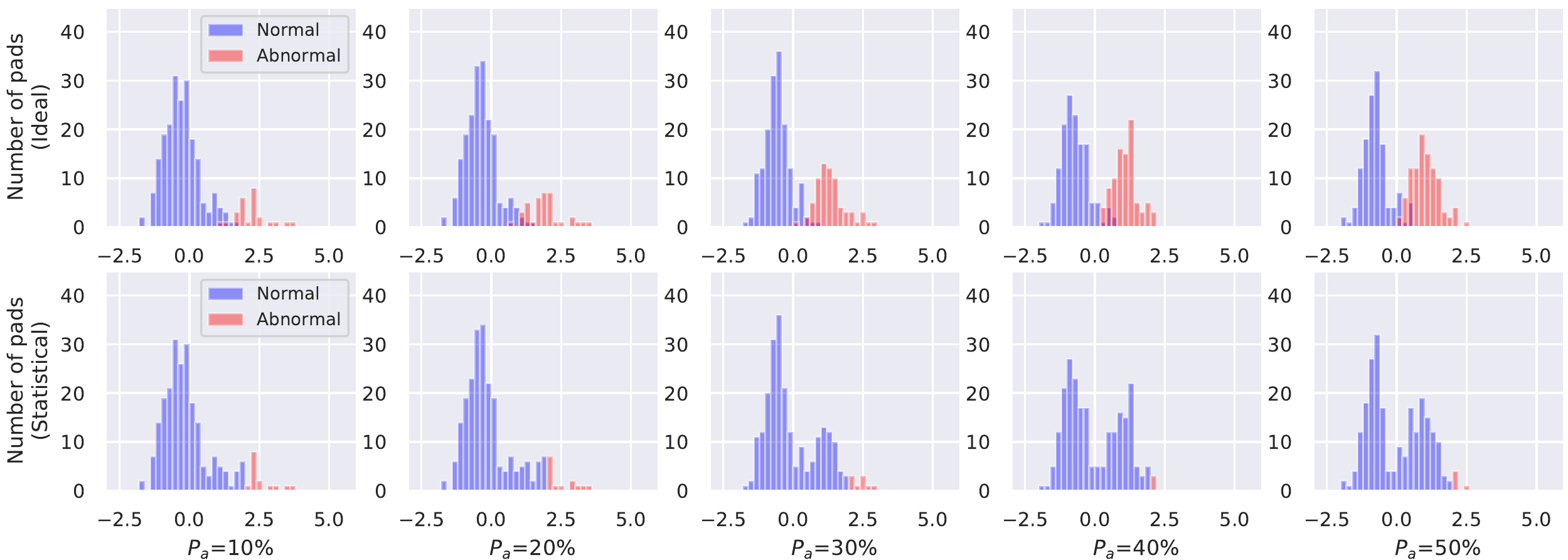}
    \caption{The solder paste volume histograms of normal and abnormal pads on the ground truth (upper graphs), and on the generated output (lower graphs).}
    \label{fig:statistical_method}
\end{figure*}

\begin{figure}
\centering
\includegraphics[width=1\linewidth]{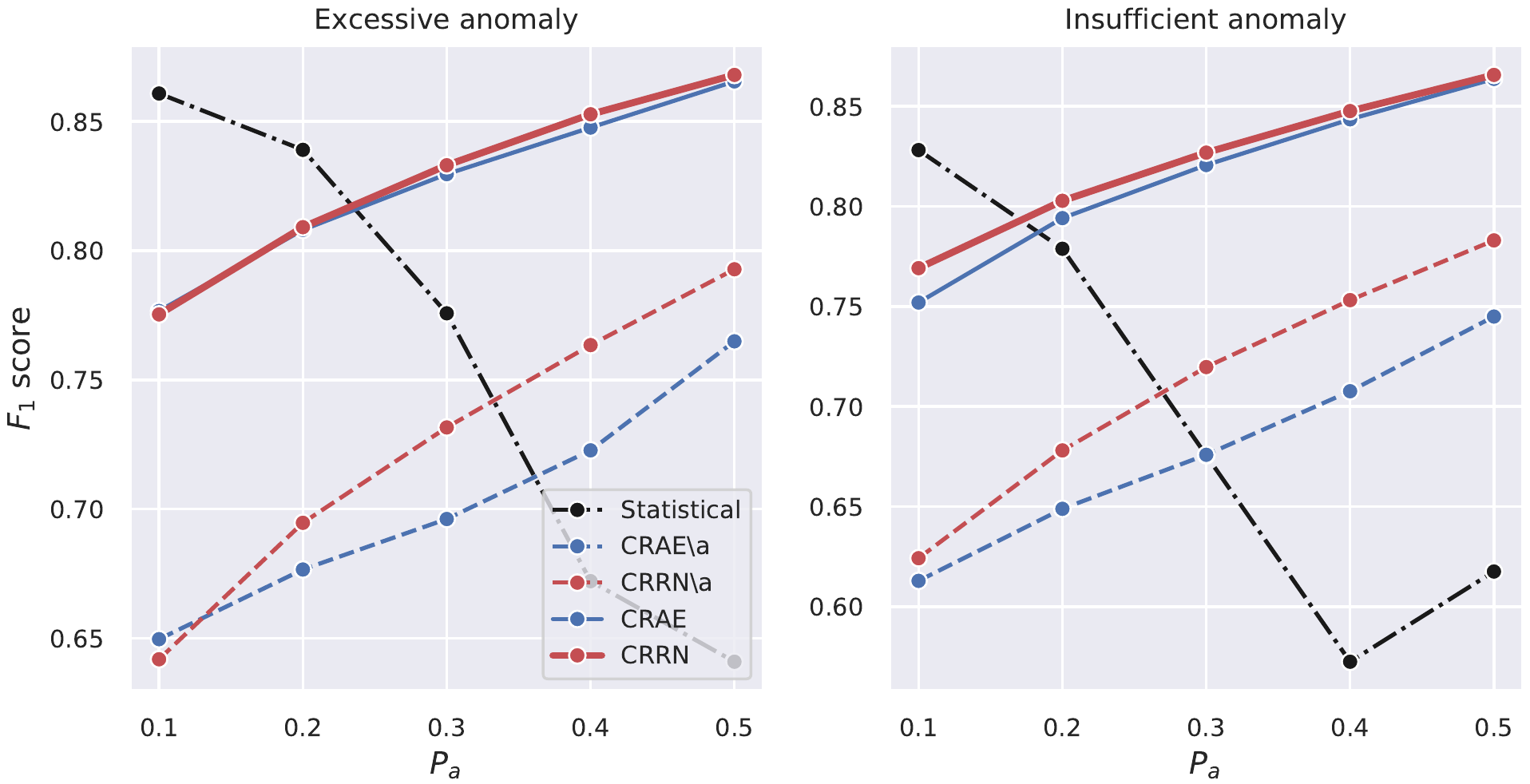}
\caption{The changes of $F_1$ score according to the changes of anomaly ratio, $P_a.$}
\label{fig:f1_score} 
\end{figure}

In the statistical method, the pads are grouped according to the shape of pads. Fig. \ref{fig:statistical_method} shows the volume histogram of the pads in one of the groups. The upper and lower graphs in the figure indicate the histograms of the ground truth and the generated output, respectively. If the anomaly ratio is relatively small, the volumes of the anomaly pads are in the tails of the entire volume distribution. However, as the anomaly ratio increases, the anomaly pads, which were minority of the entire volume, become dominant and the Gaussian distribution shifts toward the anomaly pads. Statistical method assumes that normal pads are dominant so that the pads located at the tails of the Gaussian distribution are regarded as the anomaly pads, which causes performance degradation as the anomaly ratio increases. On the other hand, the deep-learning based methods not only are superior to the statistical method, but also show the improved performance. Among the deep-learning based methods, the models with the ST-Attention mechanism are better than the models without the ST-Attention mechanism. Moreover, the accuracy of the proposed CRRN is higher than that of CRAE. The results of Experiment 1 are summarized in Fig. \ref{fig:f1_score} and Table 1. 

\subsection{Experiment 2} 
\begin{table}[]
\small
\caption{$F_1$ scores according to anomaly ratio, $P_a$. The symbol \textbackslash a indicates that the ST-Attention mechanism is not applied.}
\begin{center}
\subtable[$F_1$ score of excessive anomaly.]{
    \begin{tabular}{l?C{0.8cm}|C{0.8cm}|C{0.8cm}|C{0.8cm}|C{0.8cm}}
    \specialrule{.1em}{.05em}{.05em} 
    $P_a$ & $10\%$ & $20\%$ & $30\%$ & $40\%$ & $50\%$ \\
    \specialrule{.1em}{.05em}{.05em} 
    Statistical & \textbf{0.78} & 0.74 & 0.67 & 0.58 & 0.59 \\
    \specialrule{.05em}{.05em}{.05em} 
    CRAE\textbackslash a & 0.65 & 0.67 & 0.69 & 0.72 & 0.76 \\
    CRRN\textbackslash a & 0.64 & 0.69 & 0.73 & 0.76 & 0.79 \\
    CRAE & 0.77 & \textbf{0.80} & \textbf{0.83} & 0.84 & \textbf{0.86} \\
    CRRN & 0.77 & \textbf{0.80} & \textbf{0.83} & \textbf{0.85} & \textbf{0.86} \\
    \specialrule{.1em}{.05em}{.05em} 
    \end{tabular}
}

\subtable[$F_1$ score of insufficient anomaly.]{
    \begin{tabular}{l?C{0.8cm}|C{0.8cm}|C{0.8cm}|C{0.8cm}|C{0.8cm}}
    \specialrule{.05em}{.05em}{.05em} 
     $P_a$ & $10\%$ & $20\%$ & $30\%$ & $40\%$ & $50\%$ \\
    \specialrule{.1em}{.05em}{.05em} 
    Statistical & \textbf{0.79} & 0.75 & 0.67 & 0.57 & 0.59 \\
    \specialrule{.1em}{.05em}{.05em} 
    CRAE\textbackslash a & 0.61 & 0.64 & 0.67 & 0.70 & 0.74 \\
    CRRN\textbackslash a & 0.62 & 0.67 & 0.72 & 0.75 & 0.78 \\
    CRAE & 0.75 & 0.79 & \textbf{0.82} & \textbf{0.84} & \textbf{0.86} \\
    CRRN & 0.76 & \textbf{0.80} & \textbf{0.82} & \textbf{0.84} & \textbf{0.86} \\
    \specialrule{.1em}{.05em}{.05em} 
    \end{tabular}
}
\end{center}

\vskip -0.1in
\label{tab:space}
\end{table}

\begin{figure}
	\centering	   	
    \subfigure[]{\includegraphics[width=0.33\linewidth]{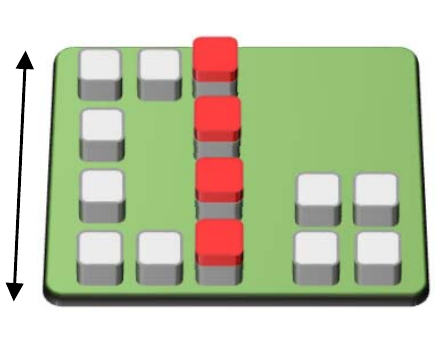}}    
   \hspace{0.1em}
    \subfigure[]{\includegraphics[width=0.29\linewidth]{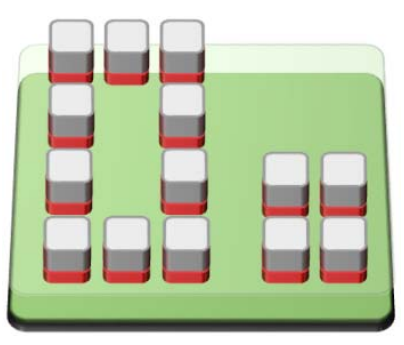}}
   \hspace{0.1em}
    \subfigure[]{\includegraphics[width=0.29\linewidth]{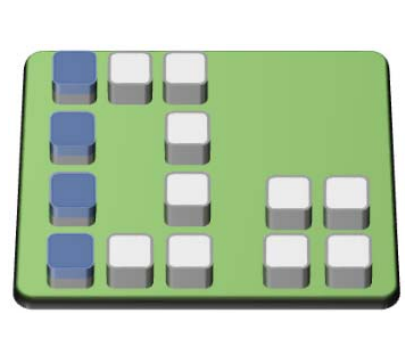}}
    \subfigure[]{\includegraphics[width=0.29\linewidth]{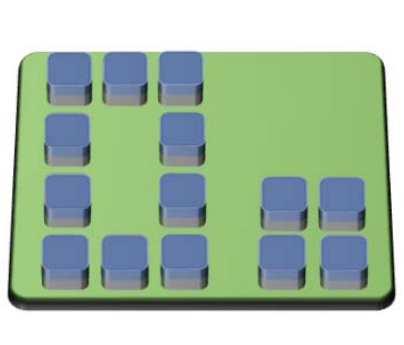}}
    \hspace{0.4em}
    \subfigure[]{\includegraphics[width=0.29\linewidth]{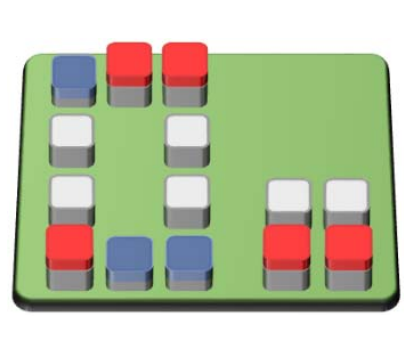}}    
    \caption{The anomaly SPI data affected by the five printer defects that cause the SPP to malfunction. (a) squeegee blade defect, (b) support defect, (c) removed area of solder paste, (d) solder no kneading, and (e) clamp defect. The arrow in (a) indicates the direction of the squeegee blade defect. In (b), the solder paste in red represents the excessively deposited solder paste as much as the board subsides due to the support defect.} 
    \label{fig:all_defects}
\end{figure}

We collected the anomaly SPI data, which were affected by the SPP defects that cause the SPP to malfunction. During solder paste printing, the squeegee blade in the SPP passes over the stencil, which deposits the solder paste in the aperture of the stencil. Fig. \ref{fig:all_defects} shows the anomaly SPI data generated by the five printer defects, which are described in detail below. 
\begin{itemize}
    \item \textbf{Squeegee blade defect}: It happens when squeegee blade has cracks. The cracked portion of the squeegee blade makes the solder paste deposited more than other regions. In the SPP, there are two types of squeegee blades with different directions, forward and backward directions. We deliberately created the squeegee blade defects only on the forward squeegee blade.
    \item \textbf{Support defect}: If the board is not maintained flat during printing, the solder paste is deposited excessively on the broad region of the board. 

    \item \textbf{Removed area of solder paste}: Repeated printing is likely to result in depositing the solder paste insufficiently. In the trail of the squeegee blade, a lot of solder paste is used for the dense parts of the apertures, so that the solder paste of these parts is insufficiently deposited. 

    \item \textbf{Solder no kneading}: Before printing, solder paste should be well kneaded with solder powder and flux. Insufficient kneading produces insufficient patterns over a large area. The solder no kneading usually occurs at the early stage of production and gradually disappears.
    
    \item \textbf{Clamp defect}: Clamp is used to fix both sides of the board. If the board is not properly secured, insufficient or excessive solder paste is deposited to the edge of the board. 
\end{itemize}

We used the collected SPI data as a benchmark dataset, which is formulated as follows:
\begin{align}
X_t^{in} & = X_t^{normal} + \epsilon_t^{generated} \nonumber \\
         & = X_t^{normal} + M_t^{label}\cdot f(t)
\label{eq:st_eq}
\end{align}
where $M_t^{label}$ is the spatial anomaly area representing anomaly pads in the SPI data. We had gradually increased the degree of the defect in the SPP over timestep. An anomaly score, $f(t)$ represents the degree of the defect respectively for squeegee blade defect, support defect, support defect, removal area of solder paste, solder no kneading, and clamp defect. $f(t)$ is approximated as a scalar value for each time step as follows: 
\begin{subequations}
\label{eq:t_eq}
\begin{align}
f(t) & = \begin{cases} 
\sigma(10t/T - 5), & \text{if } t \text{\ is odd}, \\ 
0, & \text{otherwise,} 
\end{cases}\\ 
f(t) & = \sigma(10t/T - 5), \\
f(t) & = -\sigma(10t/T - 5), \\
f(t) & = 1-\sigma(10t/T - 5), \\
f(t) & = \sigma(10t/T - 5) \text{\ or} -\sigma((10t/T - 5),
\end{align}
\end{subequations} 
\noindent where $T$ is the total number of PCB products generated. The effects of defects were not noticeable in early PCB production, and as the production progresses, the effects of defects appeared in the SPI data. Thus, we approximated the anomaly score using the sigmoid function. 

\begin{figure}
	\centering	   	
	\vspace{1em}
	\subfigure[Squeegee blade defect]{\includegraphics[width=1\linewidth]{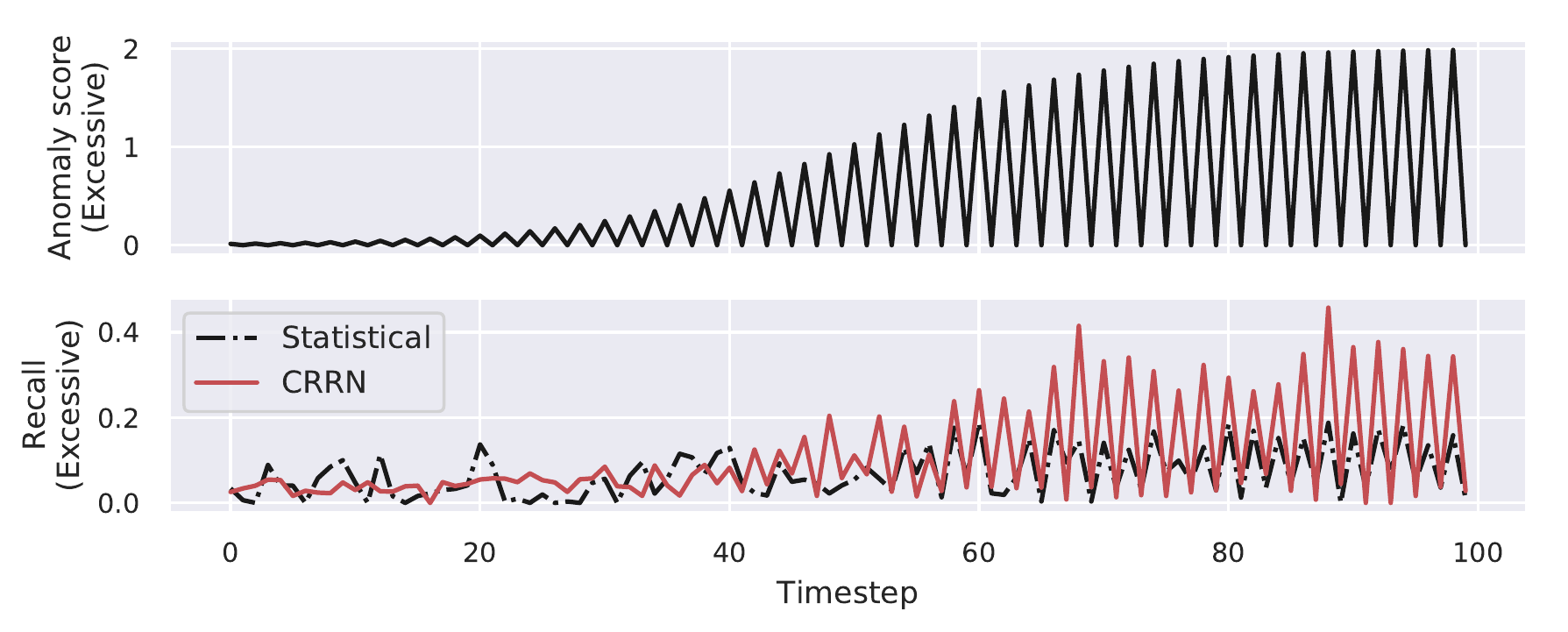}}    
    \vspace{1em}
	\subfigure[Support defect]{\includegraphics[width=1\linewidth]{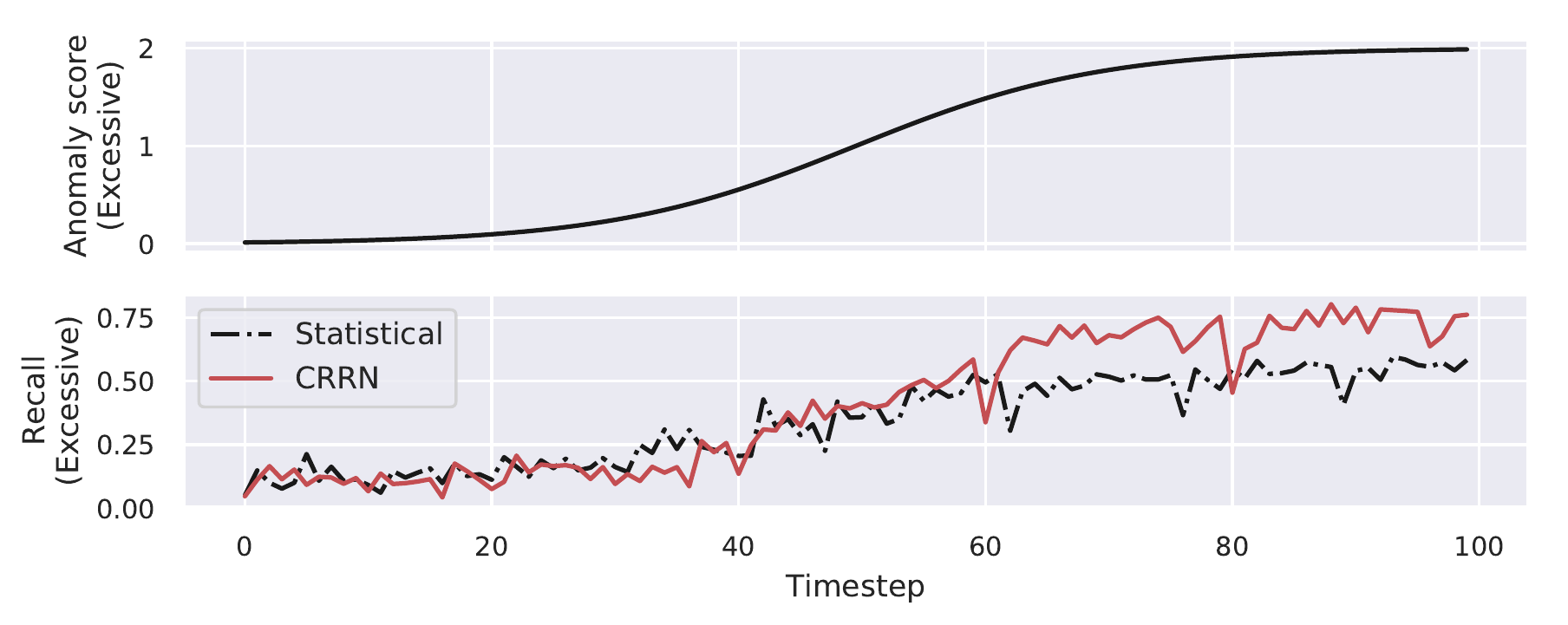}}
	\vspace{1em}
	\subfigure[Removal area of the solder paste]{\includegraphics[width=1\linewidth]{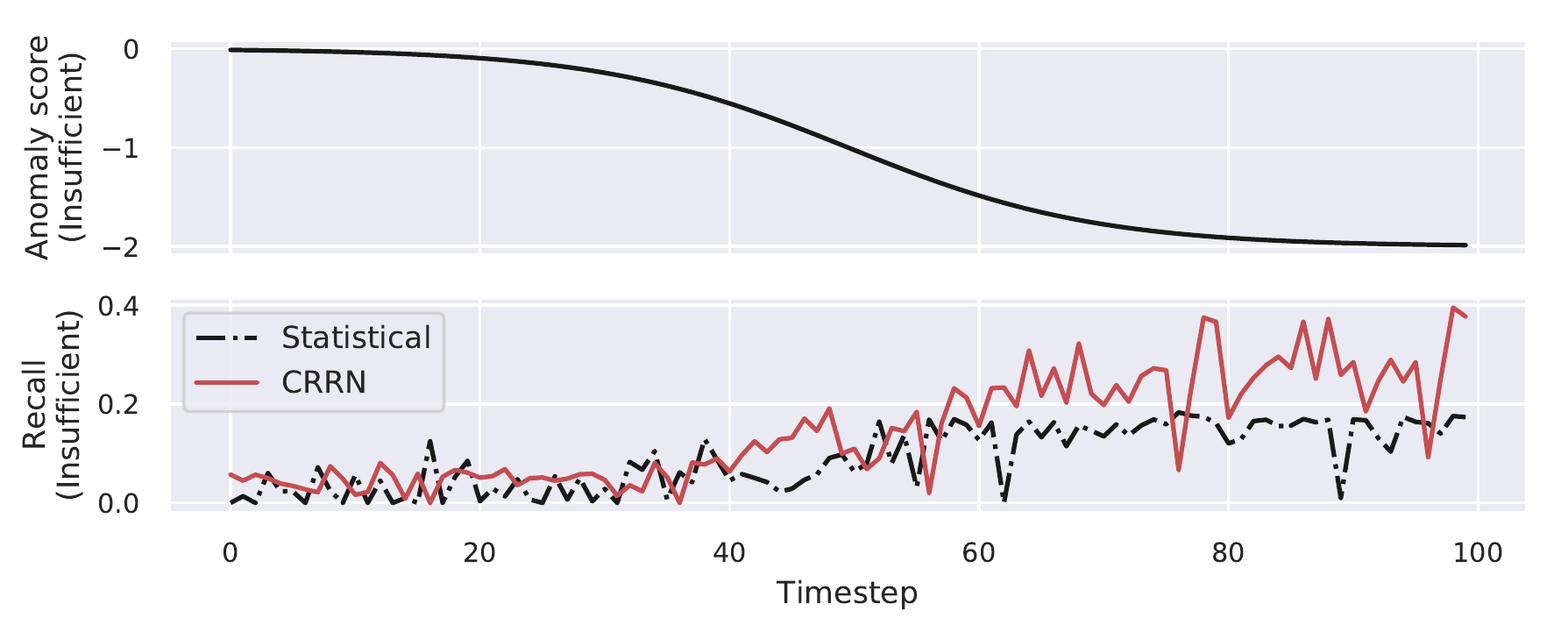}}
	\vspace{1em}
	\subfigure[Solder no kneading]{\includegraphics[width=1\linewidth]{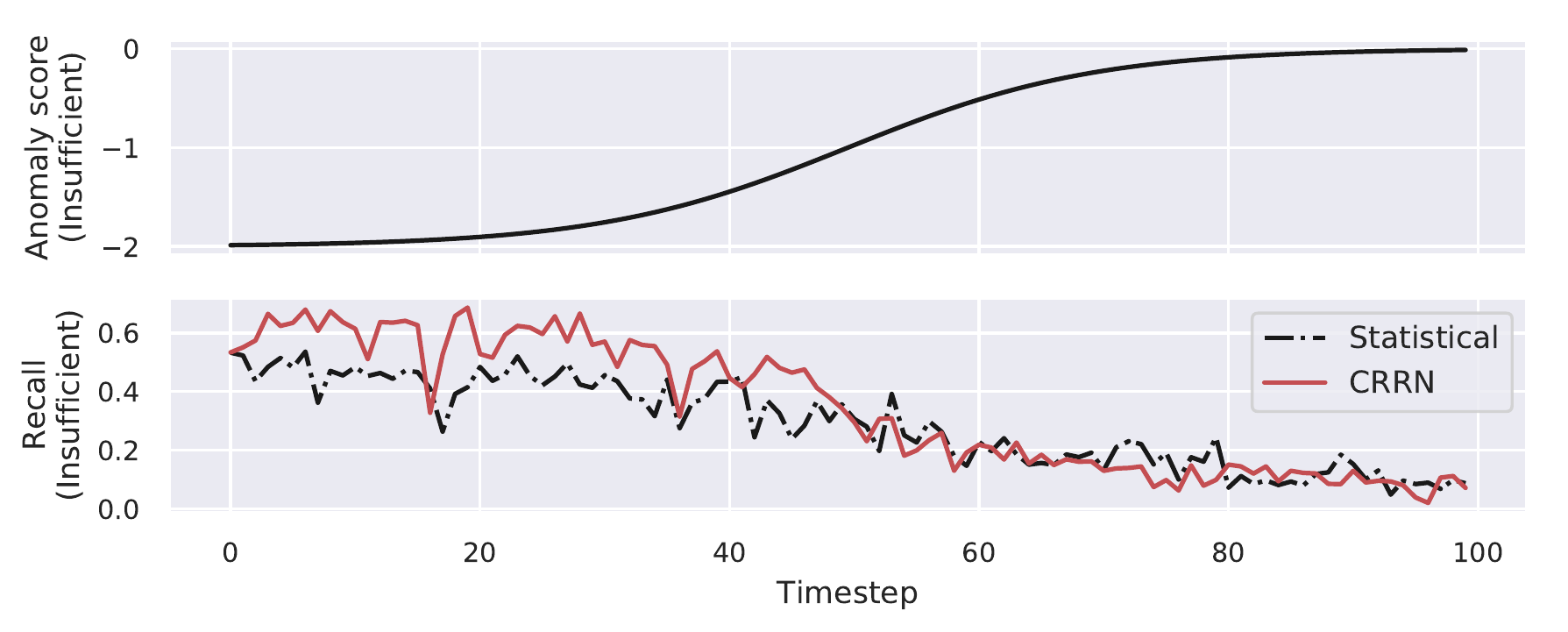}}
	\subfigure[Clamp defect]{\includegraphics[width=1\linewidth]{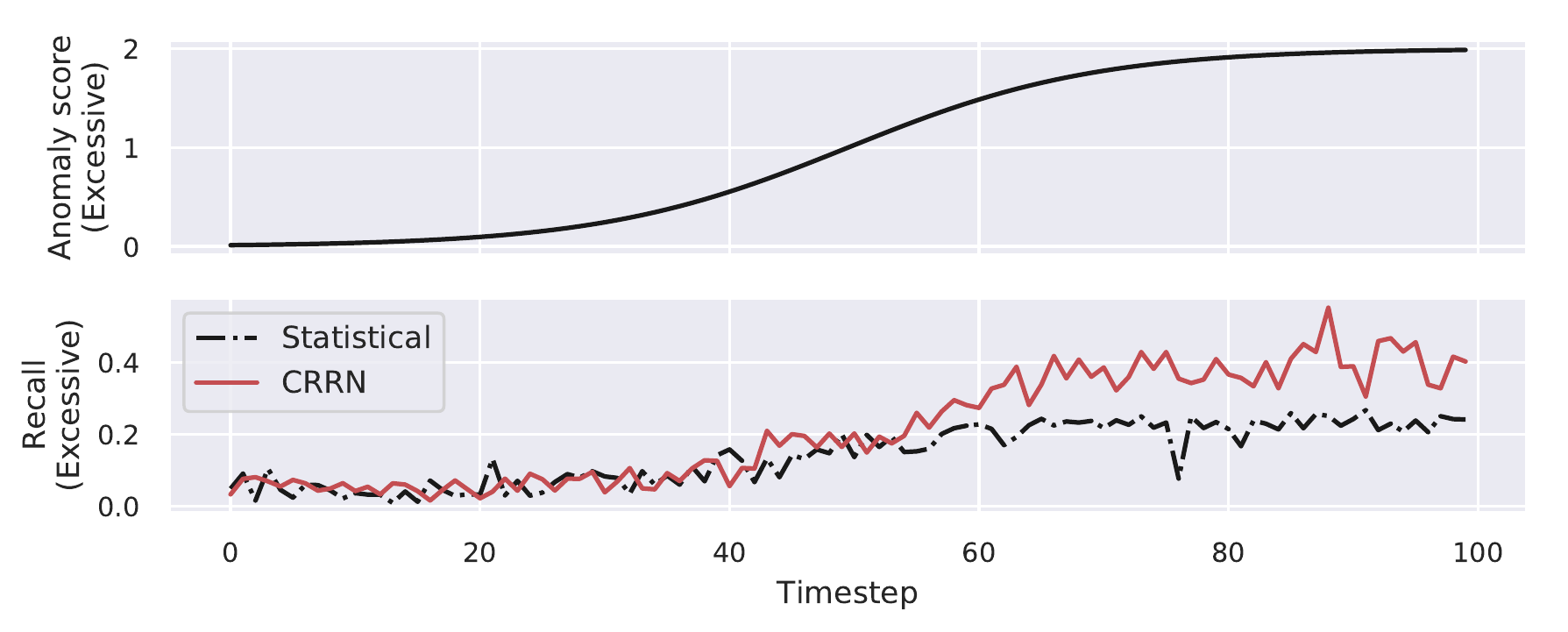}}
        \caption{Each sub-figure shows the profile of the anomaly score over timestep, $f(t)$, and the recall of each defect over timestep for each row from the top.} 
    \label{fig:defect_profile} 
\end{figure}

\begin{figure}
	\centering	   	
    \subfigure[Squeegee blade defect]{\includegraphics[width=0.87\linewidth]{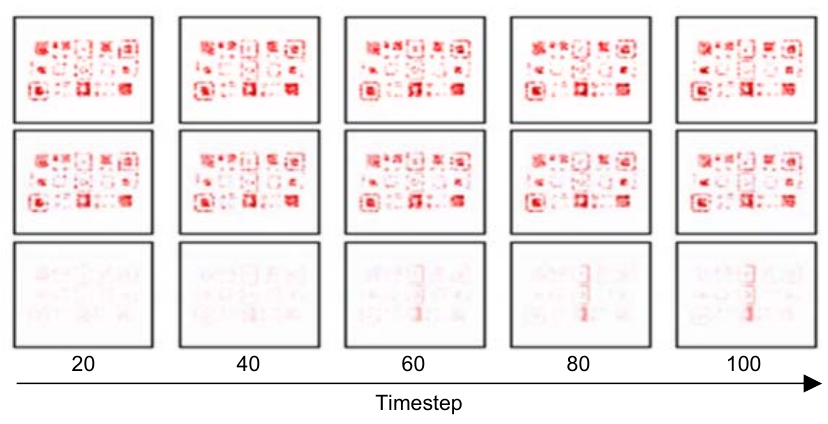}}    
    \subfigure[Support defect]{\includegraphics[width=0.87\linewidth]{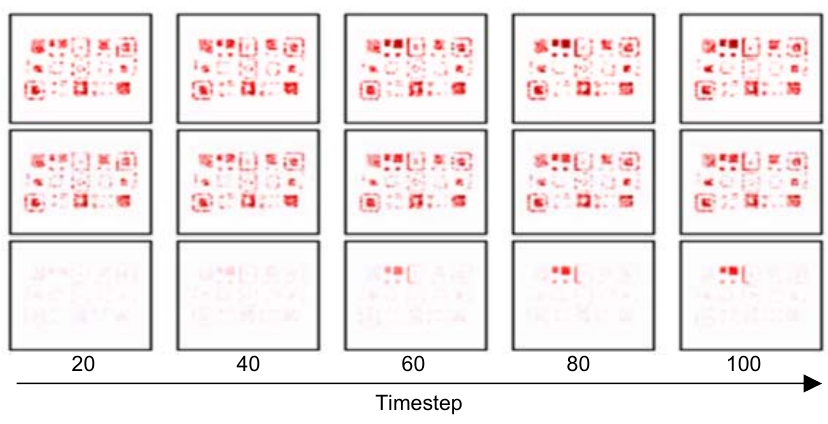}}
    \subfigure[Removal area of the solder paste]{\includegraphics[width=0.87\linewidth]{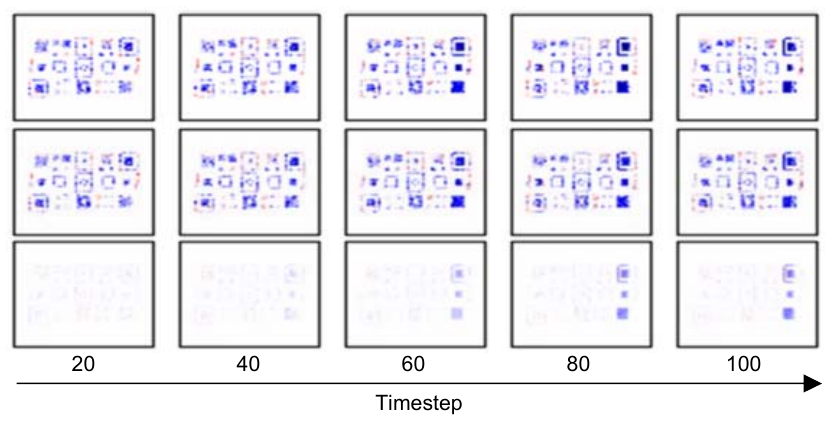}}
    \subfigure[Solder no kneading]{\includegraphics[width=0.87\linewidth]{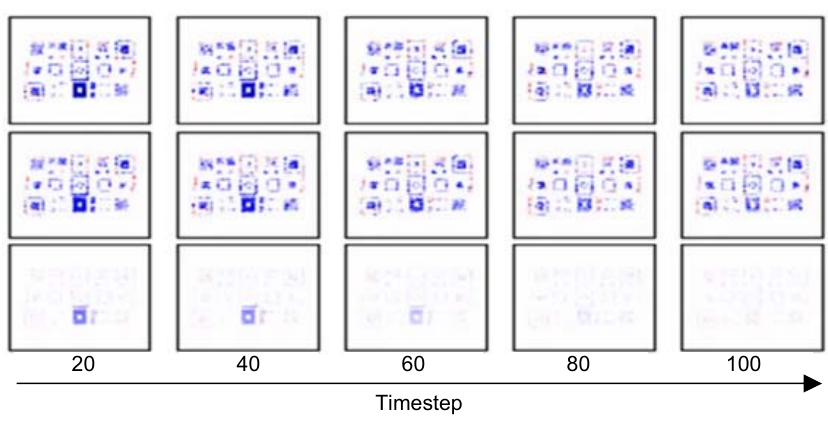}}
    \subfigure[Clamp defect]{\includegraphics[width=0.87\linewidth]{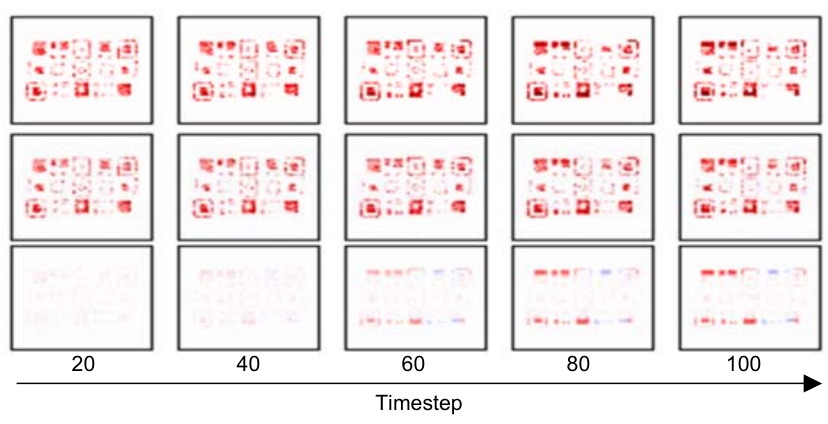}}    
    \caption{Each sub-figure shows the original SPI data, the reconstructed outputs, and the decomposed anomaly map obtained by subtracting the reconstructed outputs from the original SPI data for each row from the top.} 
    \label{fig:defect_profile2} 
\end{figure}

Fig. \ref{fig:defect_profile} displays the profile of the anomaly score, $f(t)$ for each defect type over timestep (upper graph of each subfigure). Using the selected threshold of the highest $F_1$ score obtained from Experiment 1, we evaluated the recall for each timestep (lower graph of each subfigure). We compared CRRN with the statistical method. The larger the anomaly score, the easier the detection is, because the volumes of the anomaly pads become significantly different from those of the normal pads. Thus, the recall ratio increases according to the increase of the anomaly score. In the statistical model, however, despite anomaly score increases, the recall tends to be saturated. This phenomenon is caused by the shift of the Gaussian distribution as described in Experiment 1. On the other hand,  the recall of CRRN tends to increase as the anomaly score increases. Since CRRN judges the anomaly by considering volumes of the whole pads, the performance is superior to the statistical method that judges the anomaly by using volumes of pads belonging to each group. Fig. \ref{fig:defect_profile2} shows the results of anomaly map decomposition for each defect, and each sub-figure shows the original SPI data, the reconstructed outputs, and the reconstruction errors for each row from the top. The anomaly pattern in the anomaly map becomes clear with the increase of the anomaly score. 


\subsection{Experiment 3} 
\begin{figure} 
    \subfigure[]{\includegraphics[width=0.32\linewidth]{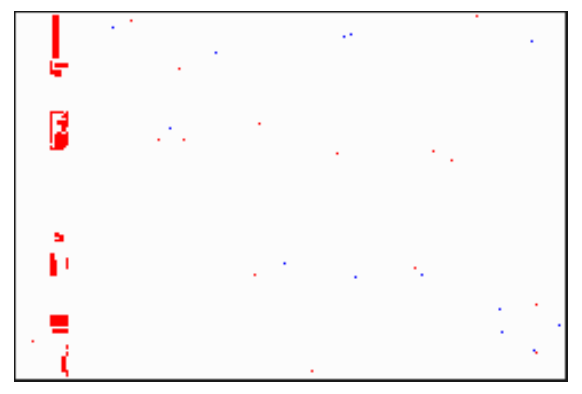}}    
	\subfigure[]{\includegraphics[width=0.32\linewidth]{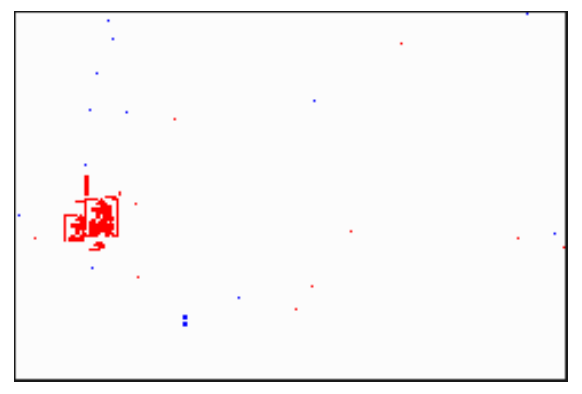}}
	\subfigure[]{\includegraphics[width=0.32\linewidth]{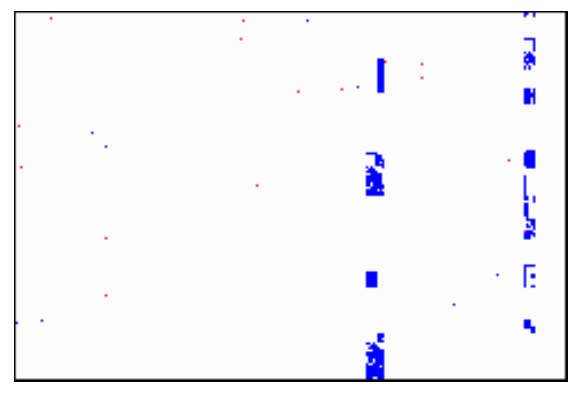}}
	\subfigure[]{\includegraphics[width=0.32\linewidth]{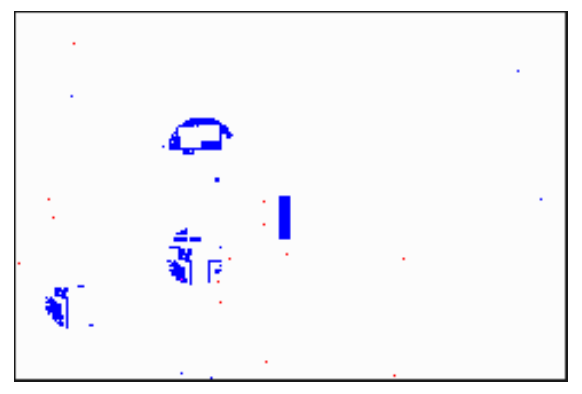}}
    \subfigure[]{\includegraphics[width=0.32\linewidth]{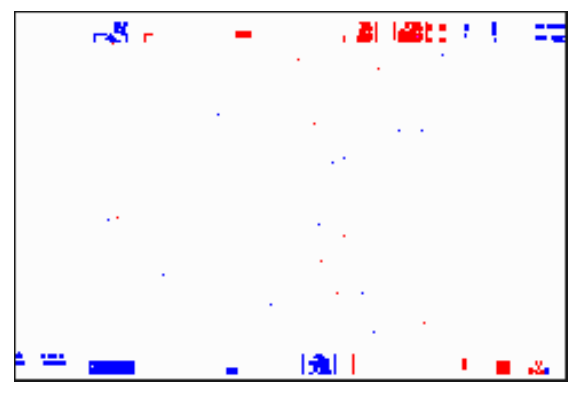}}
    \caption{Binary anomaly maps of (a) squeegee blade defect, (b) support defect, (c) removal area of solder paste, (d) solder no kneading, and (e) clamp defect.} 
    \label{fig:binary} 
\end{figure} 
The anomaly maps decomposed through CRRN can be used as features for classifying the SPP defects. We performed the defect classification task regarding the five defects introduced in Experiment 2. The decomposed anomaly map was splited into two channels, excessive and insufficient channels. The excessive channel was obtained by max pooling to preserve excessive patterns. On the other hand, insufficient channel was obtained through min pooling to preserve the insufficient pattern. Each channel was binarized using the decision threshold used in Experiment 1. Fig. \ref {fig:binary} shows the binarized channel for each defect where the excessive channel and insufficient channel are marked in red and blue, respectively. Since the SPP could have multiple defects, a combination of defect patterns may exist in the anomaly map. 
\begin{figure}
    \includegraphics[width=1.02\linewidth]{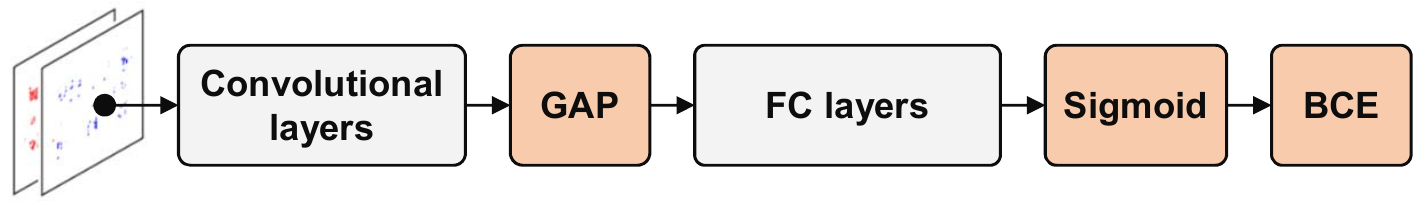}
    \caption{SPP defect classification model.}
    \label{fig:cnn}
\end{figure}

Fig. \ref{fig:cnn} shows a CNN model for the defect classification. The CNN model consists of convolutional layers for feature extraction and fully-connected layers for classification. In the convolutional layers, a pretrained Resnet-18 or Inception-v4 model was loaded and fine-tuned. Since the size of the SPI data depends on the type of the PCB, we applied a global average pooling (GAP) \cite{lin2013network} between the convolutional layers and the fully-connected layers. The number of neurons in the input layer of the fully connected layers was 512 for Resnet-18, and 1,536 for Inception-v4. In both models, the numbers of neurons in the hidden layer and in the output layer were set to 100 and 5, respectively. For multi-label classification, the output of the fully connected layer passed through a sigmoid function and the binary cross entropy was adopted as the loss function. 

We compared the performances of the classification models based on Resnet-18 and Inception-v4. The mean average precision (mAP) and the exact match ratio (EMR) were used as performance metrics. The mAP is the mean of the average precision for each defect. EMR is the ratio of the case where the target and the output are completely matched. In the case of mAP, Resnet-18 showed $91.3\%$, and Inception-v4 showed $93.8\%$. In the case of EMR, Resnet-18 showed $71.7\%$ and Inception-v4 showed $74.8\%$. The accuracy of Inception-v4 was generally superior to that of Resnet-18. Through the experiments, it was verified that anomaly maps decomposed through CRRN can be used as features for classifying defects. Note that the experiment was intended to verify the discriminative power of the anomaly map, and thus further work related to the classification model can improve the accuracy of the classification.

\section{Conclusion}
In this paper, we proposed the CRRN model to decompose anomaly patterns of SPI data caused by the printer defects that make the SPP malfunction. The CRRN consists of S-Encoder, ST-Encoder-Decoder, and S-Decoder. The ST-Encoder-Decoder consists of multiple CSTMs with the ST-Attention mechanism. CSTM has a spatiotemporal cell that can capture both spatial and temporal patterns. We also designed the ST-Attention mechanism to generate consistent outputs by dealing with the long-term dependency problem. The ST-Attention serves as a shortcut path for connecting the encoder and the decoder in CRRN. Since most of SPI data are normal, we trained CRRN using only normal SPI data. Using the trained CRRN, the anomaly map was decomposed from the SPI data with anomaly defects based on the reconstruction error. To verify the performance of CRRN, we compared CRRN with the statistical method and other deep learning-based methods. Through three experiments, we demonstrated the superior anomaly detection performance of CRRN to other methods. In addition, we proved that the anomaly map can be applied to the defect classification of the SPP. 

\ifCLASSOPTIONcaptionsoff
  \newpage
\fi



\bibliographystyle{IEEEtran}
\bibliography{references}
%

%








\end{document}